\documentclass[usenatbib]{mn2e}
\usepackage{apjfonts}
\usepackage{epsfig}
\usepackage{amssymb}
\usepackage{amsmath}
\usepackage{fixltx2e} 
\usepackage{ctable}
\usepackage{lscape}
\usepackage{url}

\newcommand{\mergercalcurl}{\url{http://www.cfa.harvard.edu/~phopkins/Site/mergercalc.html}}
\newcommand\plotone[1]
 {\centering \leavevmode \includegraphics[width={0.99\columnwidth}]{#1}}
\newcommand{\plotside}[1]
 {\centering \leavevmode \includegraphics[width={0.95\textwidth}]{#1}}
\newcommand{\acknowledgments}{\begin{small}\section*{Acknowledgments}\end{small}}
\newcommand\altaffilmark[1]{$^{#1}$}
\newcommand\altaffiltext[1]{$^{#1}$}
\newcommand{\etal}{et al.}

\newcommand{\msun}{M_{\sun}}
\newcommand{\lsun}{L_{\sun}}

\title[Starbursts from Ellipticals]{A New Empirical Method to Infer the 
Starburst History of the Universe from Local Galaxy Properties}

\author[Hopkins \etal]{
\parbox[t]{\textwidth}{ 
Philip F.~Hopkins,\thanks{E-mail:phopkins@astro.berkeley.edu}\altaffilmark{1} 
\&\ Lars Hernquist\altaffilmark{2}
}
\vspace*{6pt} \\
\altaffiltext{1}{Department of Astronomy and Theoretical Astrophysics Center, University of California Berkeley, Berkeley, CA 94720} \\
\altaffiltext{2}{Harvard-Smithsonian Center for Astrophysics, 60 
Garden Street, Cambridge, MA 02138, USA}  
}

\date{Submitted to MNRAS, September 17, 2009}
\begin{document}
\maketitle
\label{firstpage}

\begin{abstract}

The centers of ellipticals and bulges are formed dissipationally, via
gas inflows over short timescales -- the ``starburst'' mode of star
formation.  Recent work has shown that the surface brightness
profiles, kinematics, and stellar populations of spheroids can be used
to separate the dissipational component from the dissipationless
``envelope'' made up of stars formed over more extended histories in
separate objects, and violently assembled in mergers.  Given
high-resolution, detailed observations of these ``burst relic''
components of ellipticals (specifically their stellar mass surface
density profiles), together with the simple assumptions that some form
of the Kennicutt-Schmidt law holds and that the burst was indeed a
dissipational, gas-rich event, we show that it is possible to invert
the observed profiles and obtain the time and space-dependent star
formation history of each burst.  We perform this exercise using a
large sample of well-studied spheroids, which have also been used to
calibrate estimates of the ``burst relic'' populations. We show that
the implied bursts scale in magnitude, mass, and peak star formation
rate with galaxy mass in a simple manner, and provide fits for these
correlations.  The typical burst mass $M_{\rm burst}$ is $\sim10\%$ of
the total spheroid mass; the characteristic starburst timescale
implied is a nearly galaxy-mass independent $t_{\rm
burst}\sim10^{8}\,$yr; the peak SFR of the burst is $\sim M_{\rm
burst}/t_{\rm burst}$; and bursts decay subsequently in power-law
fashion as $\dot{M}_{\ast}\propto t^{-2.4}$.  As a function of time,
we obtain the spatial size of the starburst; burst sizes at peak activity
scale with burst mass in a manner similar to the observed spheroid
size-mass relation, but are smaller than the full galaxy size by a
factor $\sim10$; the size grows in time as the central, most dense
regions are more quickly depleted by star formation as $R_{\rm
burst}\propto t^{0.5}$.  Combined with observational measurements of
the nuclear stellar population ages of these systems -- i.e.\ the
distribution of times when these bursts occurred -- it is possible to
re-construct the dissipational burst contribution to the distribution
of SFRs and IR luminosity functions and luminosity density of the
Universe.  We do so, and show that these burst luminosity functions
agree well with the observed IR LFs at the brightest luminosities, at
redshifts $z\sim0-2$. At low luminosities, however, bursts are always
unimportant; the transition luminosity between these regimes increases
with redshift from the ULIRG threshold at $z\sim0$ to Hyper-LIRG
thresholds at $z\sim2$.  At the highest redshifts $z\gtrsim2$, we can
set strict upper limits on starburst magnitudes, based on the maximum
stellar mass remaining at high densities at $z=0$, and find tension
between these and estimated number counts of sub-millimeter galaxies,
implying that some change in bolometric corrections, the number counts
themselves, or the stellar IMF may be necessary.  At all redshifts,
bursts are a small fraction of the total SFR or luminosity density,
approximately $\sim5-10\%$, in good agreement with estimates of the
contribution of merger-induced star formation.

\end{abstract}


\section{Introduction}
\label{sec:intro}

Understanding the global star-formation history of the Universe
remains an important unresolved goal in cosmology.  Of particular
interest is the role played by mergers in driving star formation
and/or the infrared luminosities of massive systems.  A wide range of
observations support the view that violent, dissipational events
(e.g.\ gas-rich mergers) are important to galaxy evolution, 
and in particular that the central, dense portions of galaxy bulges and 
spheroids must be formed in such events; but less
clear is their contribution to the global star formation process.  

In the local Universe, the population of star-forming galaxies appears to
transition from ``quiescent'' (undisturbed) disks -- which dominate
the {\em total} star formation rate/IR luminosity density -- at the
luminous infrared galaxy (LIRG) threshold $10^{11}\,\lsun$
($\dot{M}_{\ast}\sim 10-20\,\msun\,{\rm yr^{-1}}$) to systems that are
clearly merging and violently disturbed at a few times this
luminosity. The most intense starbursts at $z=0$, ultraluminous
infrared galaxies (ULIRGs; $L_{\rm IR}>10^{12}\,\lsun$), are
invariably associated with mergers
\citep[e.g.][]{joseph85,sanders96:ulirgs.mergers}, with dense gas in
their centers providing material to feed black hole growth and to
boost the concentration and central phase space density of merging
spirals to match those of ellipticals
\citep{hernquist:phasespace,robertson:fp}.  Various studies have shown
that the mass involved in these starburst events is critical to
explain the relations between spirals, mergers, and ellipticals, and
has a dramatic impact on the properties of merger remnants
\citep[e.g.,][]{LakeDressler86,Doyon94,ShierFischer98,James99,
Genzel01,tacconi:ulirgs.sb.profiles,dasyra:mass.ratio.conditions,dasyra:pg.qso.dynamics,
rj:profiles,rothberg.joseph:kinematics,hopkins:cusps.ell,hopkins:cores}.

At high redshifts, bright systems dominate more and more of the IR
luminosity function
\citep[e.g.][]{lefloch:ir.lfs,perezgonzalez:ir.lfs,caputi:ir.lfs,magnelli:z1.ir.lfs}.
Merger rates increase rapidly \citep[by a factor $\sim10$ from
$z=0-2$; see e.g.][and references therein]{hopkins:merger.rates},
leading to speculation that the merger rate evolution may in fact
drive the observed evolution in the cosmic SFR density, which also
rises sharply in this interval \citep[e.g.][and references
therein]{hopkinsbeacom:sfh}.  However, many LIRGs at $z\sim1$, and
possibly ULIRGs at $z\sim2$, appear to be ``normal'' galaxies, without
dramatic morphological disturbances associated with the local
starburst population or large apparent AGN contributions
\citep{yan:z2.sf.seds,sajina:pah.qso.vs.sf,
dey:2008.dog.population,melbourne:2008.dog.morph.smooth,
dasyra:highz.ulirg.imaging.not.major}.  At the same time, even more
luminous systems appear, including large numbers of Hyper-LIRG
(HyLIRG; $L_{\rm IR}>10^{13}\,\lsun$) and bright sub-millimeter
galaxies
\citep[e.g.][]{chapman:submm.lfs,younger:highz.smgs,casey:highz.ulirg.pops}.
These systems exhibit many of the traits more commonly associated with
merger-driven starbursts, including morphological disturbances, and
may be linked to the emergence of massive, quenched (non
star-forming), compact ellipticals at times as early as $z\sim2-4$
\citep{papovich:highz.sb.gal.timescales,
younger:smg.sizes,tacconi:smg.maximal.sb.sizes,
schinnerer:submm.merger.w.compact.mol.gas,
chapman:submm.halo.clustering,tacconi:smg.mgr.lifetime.to.quiescent}.

In a series of papers, \citet{hopkins:cusps.mergers,
hopkins:cusps.ell,hopkins:cores} (hereafter H08c, H09b, H09e,
respectively), the authors combined simulation libraries of galaxy mergers with
observations of nearby ellipticals to develop a methodology to
empirically separate spheroids into their two dominant physical
components. First, a dissipationless component -- i.e.\ an
``envelope,'' formed from the violent relaxation/scattering of stars
already present in merging stellar disks that contribute to the
remnant. Because disks are extended, with low phase-space density (and
collisionless processes cannot raise this phase-space density), these
stars will necessarily dominate the profile at large radii, hence the
envelope, with a low central density.

Second, a dissipational or ``burst'' component -- i.e.\ a dense stellar relic
formed from gas which lost its angular momentum and fell into the
nucleus of the remnant, turning into stars in a compact central
starburst like those in e.g.\ local ULIRGs (although these represent
the most extreme cases).  Because gas radiates, it can collapse to
very high densities, and stars formed in a starburst will dominate the
profile within radii $\sim0.5-1\,$kpc, accounting for the high central
densities of ellipticals.  The gas will reflect that brought in from
merging disks, but could in principle also be augmented by additional
cooling or stellar mass loss in the elliptical \citep[see
e.g.][]{ciottiostriker:recycling}.  

In subsequent mergers, these two stellar components will act
dissipationlessly, but the segregation between the two is sufficient
that they remain distinct even after multiple, major ``dry''
re-mergers: i.e.\ one can still, in principle, distinguish the dense
central stellar component that is the remnant of the combined
dissipational starburst(s) from the less dense outer envelope that is
the remnant from low-density disk stars.

In H08c,  the methodology for empirically separating these
components in observed systems is presented, and tested this on samples of nearby
merger remnants from \citet{rj:profiles}.  
Comparison with other, independent constraints such as stellar population synthesis 
models \citep{titus:ssp.decomp,
reichardt:ssp.decomp,michard:ssp.decomp} 
and galaxy abundance profiles \citep{foster:metallicity.gradients.prep,
mcdermid:sauron.profiles,sanchezblazquez:ssp.gradients}, as well as direct comparison of 
simulations with observed surface brightness profiles, galaxy shapes,
and kinematics are used to demonstrate that the approach can reliably
extract the dissipational component of the galaxy (see
\S~\ref{sec:methods:obs}).  

If this general scenario is correct, then the empirically 
identified burst relic components in local spheroids represent a 
novel and powerful new constraint on the history and nature of 
dissipational starbursts. 
In this paper, we present and develop these constraints -- and 
in particular, we show that they are not merely integral constraints 
on the masses and sizes of bursts. 
We show that the unique nature of dissipational star formation, together 
with the existence of some Kennicutt-Schmidt type 
relation between gas surface densities and star formation rates, 
means that the relic profiles can be {\em inverted} to 
obtain the full time and radius-dependent star formation history 
of each galaxy starburst (\S~\ref{sec:methods}). 
We present the resulting, derived burst 
star formation histories for samples of hundreds 
of local, well-observed galaxy spheroids. 
We show that such starbursts follow broadly similar 
time-dependent behavior, with characteristic 
starburst star formation rates, durations, rise and decay rates, 
and spatial sizes that scale with galaxy mass and other properties 
according to simple scaling laws across $\sim5$ decades 
in starburst and spheroid mass (\S~\ref{sec:results:properties}). 
Combining these inferred star formation histories (and resulting 
burst lightcurves) with empirical determinations of 
the ages of each starburst, we can reconstruct the 
starburst history of the Universe, including e.g.\ the luminosity 
functions of such dissipational bursts at all redshifts, 
and their contribution to the global star formation rate 
density (\S~\ref{sec:results:history}). 
We discuss the uncertainties in this approach, 
compare with the previous, independent attempts to constrain 
these quantities via other observational methods, 
and summarize our results in \S~\ref{sec:discussion}. 

Throughout, we adopt a $\Omega_{\rm M}=0.3$, $\Omega_{\Lambda}=0.7$,
$h=0.7$ cosmology and a \citet{chabrier:imf} stellar IMF, but these
choices do not affect our conclusions.

\section{The Observations and Methodology}
\label{sec:methods}

\subsection{The Observations}
\label{sec:methods:obs}

As discussed in \S~\ref{sec:intro}, spheroid mass profiles can be
decomposed into a central relic starburst and an outer stellar
envelope.  The two components are distinct physically in the sense
that the amount and distribution of the two are determined by
dissipational and dissipationless dynamics, respectively.  In H08c,
H09b,e, the authors compile large samples of ellipticals from the studies of
\citet{jk:profiles} and \citet{lauer:bimodal.profiles}, and present
decompositions for these galaxies.  We adopt these results for the
present study.

Briefly, \citet{jk:profiles} present a $V$-band Virgo elliptical
survey, based on the complete sample of Virgo galaxies down to
extremely faint systems in \citet{binggeli:vcc} \citep[the same sample
studied in][]{cote:virgo,ferrarese:profiles}.  \citet{jk:profiles}
combine observations from a large number of sources
\citep[including][]{bender:data,caon90,caon:profiles,davis:85,kormendy:05,
lauer:85,lauer:95,lauer:centers,liu:05,peletier:profiles} and new
photometry from the McDonald Observatory, the HST archive, and the
SDSS for each of their objects which, after careful conversion to a
single photometric standard, enables accurate surface brightness
measurements over a wide dynamic range (with an estimated zero-point
accuracy of $\pm0.04\,V\,{\rm mag\, arcsec^{-2}}$).  Typically, the
galaxies in this sample have profiles spanning $\sim12-15$ magnitudes
in surface brightness, corresponding to a range of nearly four orders
of magnitude in physical radii from $\sim10\,$pc to $\sim100\,$kpc,
permitting the best simultaneous constraints on the shapes of both the
outer and inner profiles of any of the objects we study.
Unfortunately, since this is restricted to Virgo ellipticals, the
number of galaxies is limited, especially at the intermediate and high
end of the mass function.

We therefore include surface brightness profiles from
\citet{lauer:bimodal.profiles}, further supplemented by
\citet{bender:data}.  \citet{lauer:bimodal.profiles} compile $V$-band
measurements of a large number of nearby systems for which HST imaging
of galactic nuclei is available.  These include the
\citet{lauer:centers} WFPC2 data-set, the \citet{laine:03} WFPC2 BCG
sample (in which the objects are specifically selected as brightest
cluster galaxies from \citet{postmanlauer:95}), and the
\citet{lauer:95} and \citet{faber:ell.centers} WFPC1 compilations.
Details of the treatment of the profiles and conversion to a single
standard are given in \citet{lauer:bimodal.profiles}.  HST images are
combined with ground-based measurements (see references above) to
construct profiles that typically span physical radii from
$\sim10\,$pc to $\sim10-20$\,kpc.  The sample includes ellipticals
over a wide range of luminosities, down to $M_{B}\sim-15$, but is
dominated by intermediate and giant ellipticals and S0 galaxies, with
typical magnitudes $M_{B} \lesssim -18$.

H08c and H09b,e apply various tests to demonstrate that the profile
decompositions are robust. Systematic comparison with large suites of
hydrodynamic simulations as calibrators implies that the derived burst
masses and radial profiles are accurate to within a factor of $\sim2$.
Independent, purely empirical analyses of these profiles
\citep{jk:profiles} and those of similar objects
\citep[e.g.][]{ferrarese:profiles, balcells:bulge.xl} yield similar
conclusions. For most of the objects in our data-set, the profiles
include e.g.\ ellipticity, $a_{4}/a$, and $g-z$ colors as a function
of radius, as well as kinematic information.  In \citet{jk:profiles}
and \citet{hopkins:cusps.ell,hopkins:cores}, the authors show that these
secondary properties exhibit transitions that mirror the fitted
decompositions (e.g.\ transitions to diskier, more rotationally
supported, younger inner components).

Likewise, comparison with detailed resolved stellar population
properties -- i.e.\ independent constraints from stellar 
population synthesis models, abundance gradients, and 
galaxy colors -- yields good agreement where available
\citep{mcdermid:sauron.profiles,sanchezblazquez:ssp.gradients,
reda:ssp.gradients,foster:metallicity.gradients.prep,
schweizer96,titus:ssp.decomp,schweizer:7252,
schweizer:ngc34.disk,reichardt:ssp.decomp,michard:ssp.decomp}.
Moreover, in young merger remnants, where stellar populations and
colors more clearly indicate the post-starburst component, the
methodology works well \citep{rj:profiles,rothberg.joseph:kinematics,
rothberg.joseph:rotation,hopkins:cusps.mergers}.  In H09b,e, 
these results are also checked against profile data used in
\citet{bender:data,bender:ell.kinematics,bender:ell.kinematics.a4,
bender:velocity.structure}.  The latter are more limited in dynamic
range, but allow for construction of multi-color profiles in e.g.\
$V$, $R$, and $I$ bands, to test whether they are sensitive to the
band used and to construct e.g.\ stellar mass and $M_{\ast}/L$
profiles.

\begin{figure*}
    \centering
    \plotside{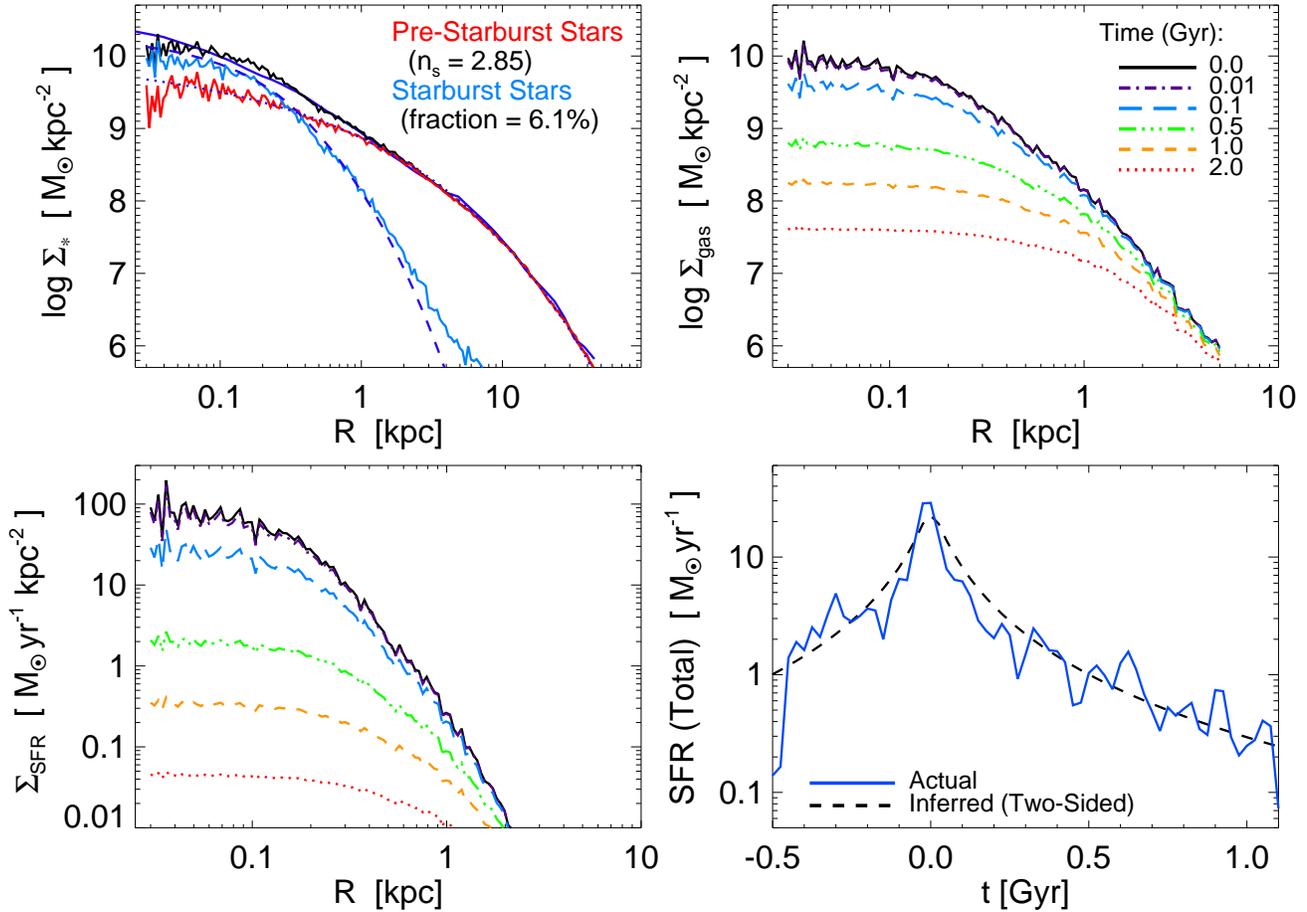}
    \caption{Illustration of the methodology used in this paper. 
    {\em Top Left:} Observed stellar mass profile of a typical $\sim L_{\ast}$ elliptical from the 
    \citet{lauer:bimodal.profiles} sample (dark blue solid). 
    We show the empirical, fitted decomposition into inner ``starburst relic'' 
    (dark blue dashed) and outer violently relaxed (dark blue dotted)
    components, from H09b,e.
    We compare a simulation which yields a remnant with similar 
    profile, kinematics, and stellar populations (total mass profile in black). 
    For the simulation, we know the true, dissipational merger-induced 
    starburst component (light blue, with labeled mass fraction), 
    and dissipationless violently relaxed component from 
    stars formed before the final merger (red; with labeled Sersic index). 
    The fitting and simulation-matching procedures yield similar 
    burst decompositions. 
    {\em Top Right:} Implied gas surface density profile of the burst ({\em ignoring} the 
    violently relaxed component, formed over much more extended periods) as a function of 
    time. At time $t=0$ (defined as the time where the implied SFR peaks), 
    we assume the burst relic was in place mostly as gas. 
    Given the observed \citet{kennicutt98} relation, this yields a SFR surface density at each 
    location.  The gas density then declines as gas is turned to stars. 
    Integrating forward, we obtain the gas (and formed stellar) mass distribution 
    at each later time (labeled). 
    {\em Bottom Left:} Corresponding SFR surface density profile. We also 
    label the half-SFR radius at each time. 
    {\em Bottom Right:} Implied {\em total} SFR (integrating the SFR density profile 
    over radius), at each time (black dashed). The ``two-sided'' estimate assumes 
    a light curve with a symmetric rise/fall. 
    We compare the true simulation SFR versus time in the best-fit simulation. 
    The two agree well -- despite the simple nature of the model, it appears to 
    be reliable in simulations with more detailed physics, and allows us to 
    recover the star formation history of a given spheroid starburst from 
    its relic mass profile. 
    \label{fig:sim.example}}
\end{figure*}

We illustrate our methodology in Figure~\ref{fig:sim.example}.  First,
we show a typical stellar mass surface density profile of an $\sim
L_{\ast}$ elliptical galaxy, from the sample of
\cite{lauer:bimodal.profiles}.  We plot the decomposition into burst
and extended envelope (violently relaxed) components, together with
some of their integral properties (stellar mass fractions, effective
radii, and fitted profile shapes in the form of the best-fit Sersic
index).  We compare the observed system to the stellar mass density
profile of one of the hydrodynamic simulations presented in H09b.  The
simulation is determined in that paper to be similar in total mass,
profile shape, and kinematic properties to the observed system (see
Table~1 therein).  In the simulation, the true, physical starburst
component from gas that loses angular momentum and falls to the center
at coalescence is known, as is the non-starburst component from
violently relaxed stars present in the disks before the final merger.
We show the separate contributions of the two components to the
stellar mass density.  The agreement is good, lending confidence to
our empirical de-composition procedure, and allowing us to consider
the simulations as a reasonable calibration sample for our approach.

\subsection{Basic Assumptions}
\label{sec:methods:assumptions}

Taking the observationally estimated burst component, this defines a
relic surface density $\Sigma_{\rm burst}(R)$.  Of course, in general,
the star formation history that yields a given stellar surface density
is non-unique.  But, if we make three simple, physically and
observationally motivated assumptions, $\Sigma_{\rm burst}(R)$ can be
inverted into a space and time-dependent burst star formation history.

These assumptions are: 
\begin{itemize}

{\item \bf (1)} That some form of the \citet{kennicutt98} law holds
for these objects, relating their star formation rate surface density
$\dot{\Sigma}_{\ast}$ to their gas surface density $\Sigma_{\rm gas}$.

{\item \bf (2)} That the relic burst mass is dominated by of order one
most massive event, i.e.\ is not constituted from the later assembly
of many independent, well-separated bursts nor by an extended series
of independent, well-separated bursts in the same galaxy.  There may be
many small events, but most of the mass should come from a single,
dominant event.

{\item \bf (3)} That this event, at its inception, involved the
central regions being (however briefly) gas-dominated. In other words,
the burst began from a gas-dominated central density and formed
largely in situ, rather than from an extended, low-gas-density trickle
in which the gas density was never nearly as large as the total/final
stellar mass density.

\end{itemize}

In detail, these assumptions are not exactly correct (we discuss this
further in \S~\ref{sec:discussion}), but most observational
constraints indicate that deviations from them are at the factor of
$\sim$ a couple level, comparable to or less than e.g.\ most of the
systematic uncertainties in the observations to which we will compare.

Regarding assumption {\bf (1)}; observations indicate that at both low
\citep[e.g.][and references therein]{kennicutt98,
bigiel:2008.mol.kennicutt.on.sub.kpc.scales,
bothwell:2009.local.sfr.demographics} and high \citep[$z\sim2-4$,
e.g.][]{bouche:z2.kennicutt} redshifts, a Kennicutt-Schmidt star
formation law applies over a large dynamic range in stellar mass and
star formation rates (from $\sim1-3000\,\msun\,{\rm yr^{-1}}$).  These
observations have also shown that the same relation pertains to a
diverse array of objects, from dwarf galaxies, through local gas-poor
disks, to near ``quenched'' red systems, to clumpy, turbulent high
redshift disks and irregular systems, to merger-induced starbursts and
ULIRGs, and high-redshift sub-millimeter galaxies, among others.  The
extent of the uncertainty appears to be in the precise
normalization/slope of the relation (with the references above
favoring power-law slopes varying from $\sim 1.4-1.7$). We take this
uncertainty into account below, and show that it has a relatively
minor effect on our conclusions.  The \citet{kennicutt98} relation may
break down at very low surface densities, where there is some star
formation threshold, but our analysis here is specifically for very
high surface-density systems.

Regarding assumption {\bf (2)}; this is well-motivated by a
combination of theoretical models and observations of galaxy stellar
populations and structural properties. In most models, nuclear burst
masses are dominated by the largest, most recent event \citep[see
e.g.][]{khochfar:sam,croton:sam,delucia:sam,somerville:new.sam,
hopkins:merger.rates}. Mergers and other violent processes that can
drive large quantities of gas into galaxy centers are not so common as
to yield many bursts of equal strength, and even if systems had
earlier bursts at much higher redshifts, they would have grown by a
large factor in mass since those times, such that the most recent
event would dominate \citep[see references above
and][]{weinzirl:b.t.dist,hopkins:merger.rates.methods}.

Observationally, evidence from nuclear kinematics and profile shapes
supports the view that most of the mass in spheroid starburst
components, especially in the cusp ellipticals that dominate ($>90\%$)
the total mass density of spheroids and bulges, was assembled in situ
by the most recent starburst, without significant modification since
that time \citep[e.g.][]{faber:ell.centers,
kormendy99,quillen:00,rest:01,ravindranath:01,laine:03,lauer:centers,
simien:kinematics.1,simien:kinematics.2,simien:kinematics.3,
emsellem:sauron.rotation.data,emsellem:sauron.rotation,
mcdermid:sauron.profiles,cappellari:anisotropy,
lauer:bimodal.profiles,ferrarese:profiles,cote:smooth.transition,
hopkins:cusps.fp,jk:profiles}.  Even in core ellipticals, thought to
have been affected by subsequent re-mergers, the kinematics require
relatively little mixing of major populations, with of order a couple
similar-mass merged systems (i.e.\ introducing a factor $\sim2$
uncertainty) -- more minor mergers may be dynamically important, but
will (by definition) contribute little mass.  

Most important, this assumption is supported by direct observations.
which indicate that the central component is well-matched by a single
stellar population (with the background galaxy -- i.e.\ the violently
relaxed component -- being some older population), usually better so
than by modeling it is as having resulted from star formation extended
in time \citep{trager:ages,nelan05:ages,thomas05:ages,gallazzi:ssps,
gallazzi06:ages,lauer:centers,kuntschner:line.strength.maps,
mcdermid:sauron.profiles}.  The metallicities and
$\alpha$-enhancements of these regions require that the stars there
formed rapidly, in short-lived events or in events closely spaced in
time (later series of bursts, or independent well-spaced events, being
ruled out by the enrichment patterns).

Regarding assumption {\bf (3)}; the observations argue that the system
must have been (at least briefly) dominated by a strong gas inflow,
that then formed stars more or less in situ.  The short timescales of
star formation indicated by stellar populations do not allow for a
trickle of gas over an extended period of time. If assumption {\bf
(1)} applies, the corresponding star formation timescale required to
match the abundances is close to that obtained if all the gas were in
place at once, and formed stars according to the \citet{kennicutt98}
relation -- i.e.\ the shortest allowable timescale. Moreover, in
hydrodynamic simulations of galaxy-galaxy mergers or dissipational
collapse, this is almost always the case -- gas flows in on the local
free-fall time.  However, star formation is inefficient on the the
free-fall timescale (the \citet{kennicutt98} relation implying star
formation efficiencies of $\sim3-10\%$ per free-fall time), and so
catches up over a short period following the initial gas-rich phase of
inflow.  In addition, the disky, rotational kinematics distinctive of
the burst relics require in situ formation from initially gas-rich
configurations \citep[see
e.g.][]{naab:gas,cox:kinematics,hopkins:cores}.

Note that the assumptions outlined here clearly do not apply for 
stars not formed in bursts. Although
observations suggest that the \cite{kennicutt98} law applies (although
there may be a surface density threshold for star formation), it is
established, in contrast to the central, burst populations, that the
extended, violently relaxed stars {\em were} formed through star
formation histories more extended in time. Disks, of course, are fed
through continuous accretion, rather than a single massive inflow; so
both assumptions {\bf (2)} and {\bf (3)} break down. Moreover, the
outer components of ellipticals are formed via violent relaxation --
i.e.\ by definition from merging already-formed stars from other
galaxies -- and so have no reason to be a homogeneous population. In
fact, especially in massive ellipticals, it is cosmologically expected
that such portions of the galaxy are made up of debris from many small
systems \citep{gallagherostriker72,
hernquist:phasespace,delucia:ell.formation,delucia:sam,
hopkins:maximum.surface.densities,hopkins:r.z.evol,
khochfar:size.evolution.model,naab:size.evol.from.minor.mergers}.  It
is likely then, that these components -- $\sim 90\%$ of the stars, in
typical $\sim L_{\ast}$ systems -- have star formation histories that
are not trivially invertible.

\subsection{Methodology}
\label{sec:methods:method}

Together, these assumptions allow us to invert an observed 
$\Sigma_{\rm burst}(R)$. 
If the relic burst mass is dominated by of order a single major 
event, and that event was indeed a dissipational, gas-rich and/or rapid event, 
then the  gas density at the beginning of the burst (which 
we define for now, arbitrarily, as $t=0$)
should be given approximately by 
\begin{equation}
\Sigma_{\rm gas}(R,\,t=0) \approx \Sigma_{\rm burst}(R)\ .
\label{eqn:sigma.init}
\end{equation}
The \citet{kennicutt98} law relates the star formation rate surface 
density -- and hence the gas depletion rate -- to the gas surface density through 
\begin{equation}
\dot{\Sigma}_{\ast} = - \frac{{\rm d}}{{\rm d}t}{\Sigma}_{\rm gas} \approx 
1.5\times10^{-4}\,\msun\,{\rm yr^{-1}\,kpc^{-2}}
\,{\Bigl (}\frac{\Sigma_{\rm gas}}{\msun\,{\rm pc^{-2}}} {\Bigr )}^{n_{K}} 
\label{eqn:kennicutt}
\end{equation}
where the normalization comes from \citet{kennicutt98} 
corrected for our adopted \citet{chabrier:imf} IMF, here 
chosen fixed for Milky-Way like systems (where the relation is best 
calibrated). The index $n_{K}\approx1.4$ is suggested by 
observations of the local and intermediate-redshift Universe, 
and so we will adopt this as our fiducial value. However, a 
higher index of $n_{K}\approx1.7$ has been suggested from 
observations (albeit more uncertain) of systems at high redshifts
with extremely intense 
SFRs \citep{bouche:z2.kennicutt}; we therefore compare
our results to those obtained for such a higher $n_{K}$. 

Note that since we are ultimately interested in the relic star
formation rate and time-averaged gas depletion rate, stellar mass loss
can be absorbed into the normalization of the SFR-surface density
relation (in essence, adopting the instantaneous recycling
approximation). But, in general, we find that for reasonable recycling
fractions from e.g.\ the models of \citet{BC03}, the resulting
differences are less than those between the choice of the slope
$n_{K}$. Likewise, the effects of long term stellar mass loss
(although, post-burst, a typical IMF will yield just $\sim20\%$
stellar mass loss) are generally smaller than uncertainties arising
from our primary assumptions above. Fortunately, the uncertainties in
both (e.g.\ the singularity of the burst versus degree of stellar mass
loss) have opposing signs, so experimenting with reasonable variations
in each yields little net difference in our predictions.

With this relation, we can then integrate forward in time from the
initial gas density in each annulus, computing new gas and stellar
densities after some time interval, and then determining a new
$\dot{\Sigma}_{\ast}$ in that annulus.  We show the results of this
procedure for our illustrative example in
Figure~\ref{fig:sim.example}.  At the initial time, the gas density is
given by the surface density profile of the relic starburst component
isolated from the observed profile (as seen in the Figure, it makes no
difference if we take the empirical fit or starburst component derived
frmo the best-fit simulation).  As discussed above, we subtract out
the non-starburst component.  This leads to a corresponding SFR
surface density.  This depletes gas, leading to a lower gas surface
density.  The evolution is most rapid in the central regions, where
the densities are highest, because the Kennicutt-Schmidt relation is
super-linear in surface density.  At several times, we plot the
remaining gas and corresponding SFR surface densities, which both
decrease and become more extended as the dense, central regions
become depleted.

Integrating the SFR surface density over $R$ 
at each time, we obtain the total 
SFR in the burst at $t\ge0$, 
\begin{equation}
\dot{M}_{\ast} = \int_{0}^{\infty} \dot{\Sigma}_{\ast}\,2\pi\,R\,{\rm d}R \ .
\end{equation}
Moreover, if we assume that light, in the UV, radio, or IR, 
traces the star formation rate locally in some annulus, 
using e.g.\ the relation 
\begin{equation}
L_{\rm IR} \approx 1.14\times10^{10}\,\lsun\,{\Bigl (} \frac{\dot{M}_{\ast}}{\msun\,{\rm yr^{-1}}} {\Bigr)}\ 
\label{eqn:mdot.to.lir}
\end{equation}
from \citet{kennicutt98} (adjusted as appropriate for the 
\citet{chabrier:imf} IMF adopted here), 
we then we obtain the light distribution, and can 
evaluate e.g.\ the half-luminosity radius at each instant in time.
\footnote{More sophisticated, luminosity and galaxy property-dependent 
conversions have been proposed, but they largely differ at low 
IR luminosities in e.g.\ extended disks. Since the systems of interest 
here are massive starbursts, we find that the alternative conversions from e.g.\ 
\citet{buat:extinction.vs.LIR} or \citet{jonsson:sunrise.attenuation} 
make no difference to our conclusions. 
}

Figure~\ref{fig:sim.example} shows the resulting total SFR as a function of time. 
Note that the procedure above, strictly speaking, defines only a one-sided 
SFR for $t>0$ -- i.e.\ times later than some initial burst peak,
$\dot{M}_{\ast}(t > 0)$. 
Real starbursts, of course, have a rise as well as a fall after their peaks, 
and simulations (as well as simple dynamical considerations) 
indicate that -- to lowest order -- the rises and falls are 
roughly symmetric \citep{dimatteo:merger.induced.sb.sims,
cox:massratio.starbursts}. 
If we assume such symmetry, it is trivial to convert our solution for 
the burst from $t>0$ to that for all times, as follows. 

Our modeling gives a one-sided solution for the SFR versus time at
$t>0$; namely $\dot{M}_{\ast}|_{\rm one\ sided}(t > 0) \equiv f(t>0)$.
Physically, our symmetry assumption means that half the time at each
luminosity is spent on each side of the peak.  Quantitatively, for
$t<0$ (pre-peak), the SFR $\dot{M}_{\ast}(t\,|\,t<0)\rightarrow
f(-t/2)$, and for $t>0$ (post-peak),
$\dot{M}_{\ast}(t\,|\,t>0)\rightarrow f(+t/2)$. This is the symmetric
or ``two-sided'' light curve.  This choice guarantees that the {\em
total} stellar mass formed in the burst is preserved, and enforces the
symmetry constraint.  It is, therefore, appropriate for comparison to
full time-dependent light-curves.  However, we stress that for {\em
all} the quantities derived in what follows, it makes no difference.
whether we assume a one-sided or symmetric two-sided lightcurve.  The
duration of the burst is the same.  (If it is defined as some $t_{\rm
burst}$, we simply have to change our notation from referring to $t =
0$ to $t=t_{\rm burst}$, to $t = -t_{\rm burst}/2$ to $t=+t_{\rm
burst}/2$; but the zero point of time is arbitrary in any case.)
Moreover, for the purposes of luminosity functions and SFR densities,
the relevant quantity is the time spent in each SFR interval -- this
is, by definition, identical whether or not we split that time across
two sides of a peak.

We compare the integrated SFR as a function of time from our
illustrative example in Figure~\ref{fig:sim.example}, inferred from
the relic starburst profile, to the true simulation SFR as a function of time
in the full hydrodynamic simulation that produces such a profile. The
agreement is excellent, verifying our procedure.  This is despite the
fact that, in the full simulation, many of our assumptions are not
true in detail.  For example, the simulations obey a local (not
global) Kennicutt-Schmidt law; thus clumping and instabilities can
accelerate star formation locally.  The simulations also account for
feedback from stars, stellar winds, and black hole accretion, which
can remove and recycle gas. The SFR is not precisely trivial or
monotonic in time. And the gas inflows do not occur instantaneously
(so that the burst would start exactly from pure gas at $t=0$, as
assumed here).  Nevertheless, it is clear that these details lead to
almost no deviation between the true SFR versus time and that inferred
from our simple procedure, given the observed relic starburst/inner
component of the system.  This is in part because some of these
effects are small.  Also, several tend to offset one another, so that,
{\em on average}, the dominant physical considerations are the
validity of an average Kennicutt-Schmidt law and the rapid, initially
gas-rich nature of dissipational starbursts.

\section{Results}
\label{sec:results}

\subsection{Typical Behavior: Some Examples}
\label{sec:results:typical}

\begin{figure}
    \centering
    \plotone{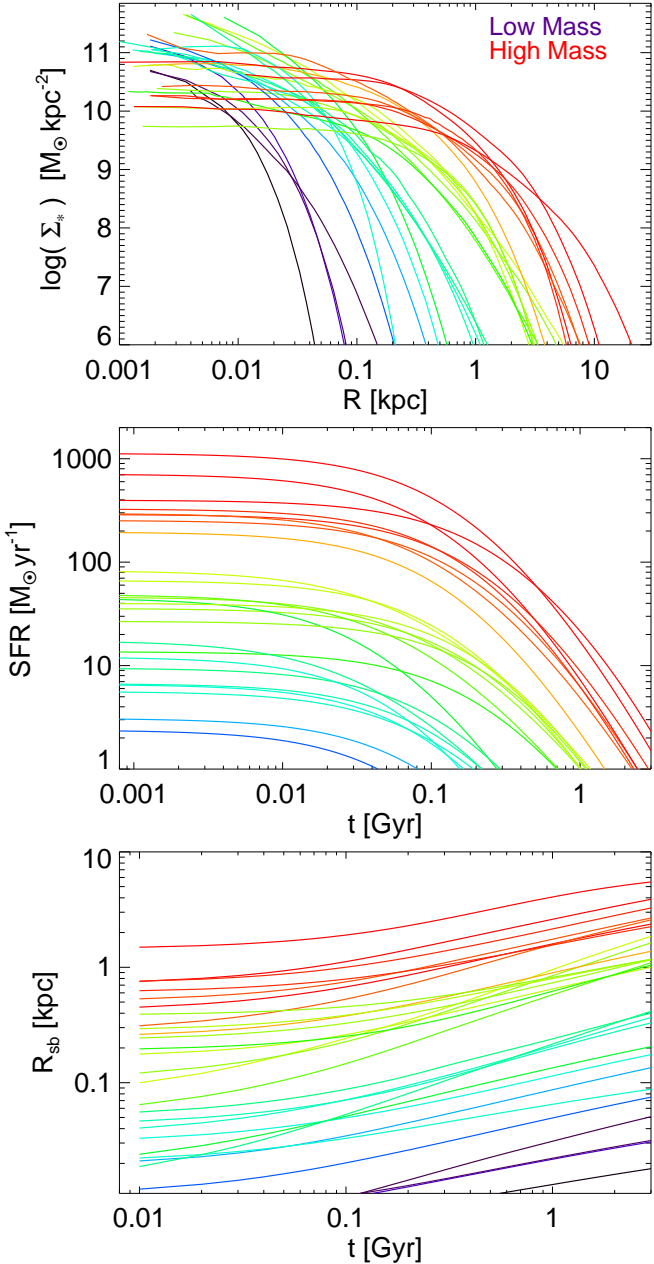}
    \caption{Typical results from recovering the burst 
    star formation histories of spheroids. 
    {\em Top:} Surface stellar mass density profiles of the empirically 
    fitted burst component alone, for each observed galaxy in the \citet{jk:profiles} 
    Virgo elliptical sample. Colors denote the total mass of each spheroid, 
    from $10^{9}\,\msun$ (violet) to $\sim10^{12}\,\msun$ (red). 
    {\em Middle:} Implied burst SFR versus time since the peak of the 
    starburst. The behavior is roughly self-similar and well-approximated 
    by Equation~\ref{eqn:mdot.time}. 
    {\em Bottom:} Burst size scale (radius enclosing $1/2$ the total star 
    formation) versus time. A similar 
    scaling applies (Equation~\ref{eqn:r.time}). 
    \label{fig:examples}}
\end{figure}

Figure~\ref{fig:examples} illustrates the results of 
our starburst inversion procedure for a 
representative sub-sample of the \citet{jk:profiles} Virgo 
ellipticals. We first show ({\em top})
the stellar mass surface density profiles of
the inner/dissipational/burst component in each 
object ({\em not} the total stellar mass density profile), as 
obtained using the two-component decomposition in 
\citet{hopkins:cusps.ell,hopkins:cores}, over the dynamic 
range covered by the observations from \citet{jk:profiles}. 
In other words, we have already subtracted off the empirically 
estimated non-burst stellar component.
For each system, using the methodology 
described in \S~\ref{sec:methods}, we convert the inferred 
surface brightness profile to a time-dependent total star formation rate, 
as illustrated in Figure~\ref{fig:sim.example}. We show 
the one-sided ($t>0$) SFR versus time so obtained, in a logarithmic scale 
(this highlights the relevant power-law behavior). 
The total burst star formation rates decay from some initial maximum in a 
power-law like fashion.  We find that good fits can 
be obtained to the time-dependent light curves with the 
functional form
\begin{equation}
\dot{M}_{\ast}(t) = \frac{\dot{M}_{\ast}(0)}{[1 + (t/t_{0})]^{\beta}}
\Longleftrightarrow 
\frac{\dot{M}_{\ast}(0)}{[1 + (|t|/[t_{0}/2])]^{\beta}}\  \ .
\label{eqn:mdot.time}
\end{equation}
The first 
is a one-sided light curve appropriate if the burst occurs only 
for $t>0$, and the second is a two-sided curve symmetric 
about $t=0$ (see \S~\ref{sec:methods}).  
The two functional forms shown are statistically 
equivalent for our purposes, and we therefore 
refer to them interchangeably. 
Fitting such a curve to each profile in Figure~\ref{fig:examples}, the 
best-fit curves are indistinguishable on the scale shown from the 
numerical calculations. 
We discuss the best-fit values below, but in general note that 
the systems tend to cluster around a power-law decay with 
$\beta\sim2.5$, and clearly exhibit a characteristic decay timescale 
of $\sim 0.1\,$Gyr. 

Note that, with such a functional form, there are straightforward relations
among several parameters. The half-life of the starburst 
$t_{1/2}$, i.e.\ the time for the starburst to decay to $1/2$ of its 
maximum luminosity/star formation rate, is simply
\begin{equation}
t_{1/2} = (2^{1/\beta} - 1)\,t_{0}\ .
\label{eqn:t12.vs.t0}
\end{equation}
Likewise, the peak SFR of the burst, $\dot{M}_{\ast}(0)$, can be 
trivially related (knowing the burst duration $t_{1/2}$ and $\beta$) 
to the total stellar mass formed in the burst, 
via the integral constraint that $M_{\rm burst} = \int^{\infty}_{0} \dot{M}_{\ast}(t)\,{\rm d}t$, 
giving
\begin{equation}
\dot{M}_{\ast}(0) = (\beta-1)\,\frac{M_{\rm burst}}{t_{0}} 
= [(\beta-1)\,(2^{1/\beta}-1)]\,\frac{M_{\rm burst}}{t_{1/2}}\ . 
\label{eqn:mdotmax}
\end{equation}
For convenience, we will use $t_{1/2}$ as our timescale of interest below, 
as it has a straightforward interpretation independent of $\beta$.
Moreover, for the $\beta$ values of interest, the coefficient 
$[(\beta-1)\,(2^{1/\beta}-1)]$ in Equation~\ref{eqn:mdotmax} 
is nearly constant at $\approx0.45-0.50$. 

At each time, we can also estimate the effective radius of the 
starburst region; we specifically define this as the half-luminosity 
radius, where we assume the surface luminosity density (in e.g.\ the infrared) 
is proportional to the gas surface density at each instant. 
We show this as well ({\em bottom}) in Figure~\ref{fig:examples}. 
In analogy to the fitted total star formation rate, we can similarly 
describe this evolution by a power-law of the form 
\begin{equation}
\dot{R}_{\rm sb}(t) = R_{\rm sb}(0){[1 + (t/t_{0,\,r})]^{\beta_{r}}}\ .
\label{eqn:r.time}
\end{equation}
The radius doubling time $t_{1/2,\,r}$ is trivially related to $t_{0,\,r}$ 
by the same relation as Equation~\ref{eqn:t12.vs.t0} above;
i.e.\ $t_{1/2,\,r} = (2^{1/\beta_{r}}-1)\,t_{0,\,r}$.

\subsection{Properties of Bursts versus Mass and Scale}
\label{sec:results:properties}

\subsubsection{Masses, Timescales, and Star Formation Rates}
\label{sec:results:properties:mass}

\begin{figure*}
    \centering
    \plotside{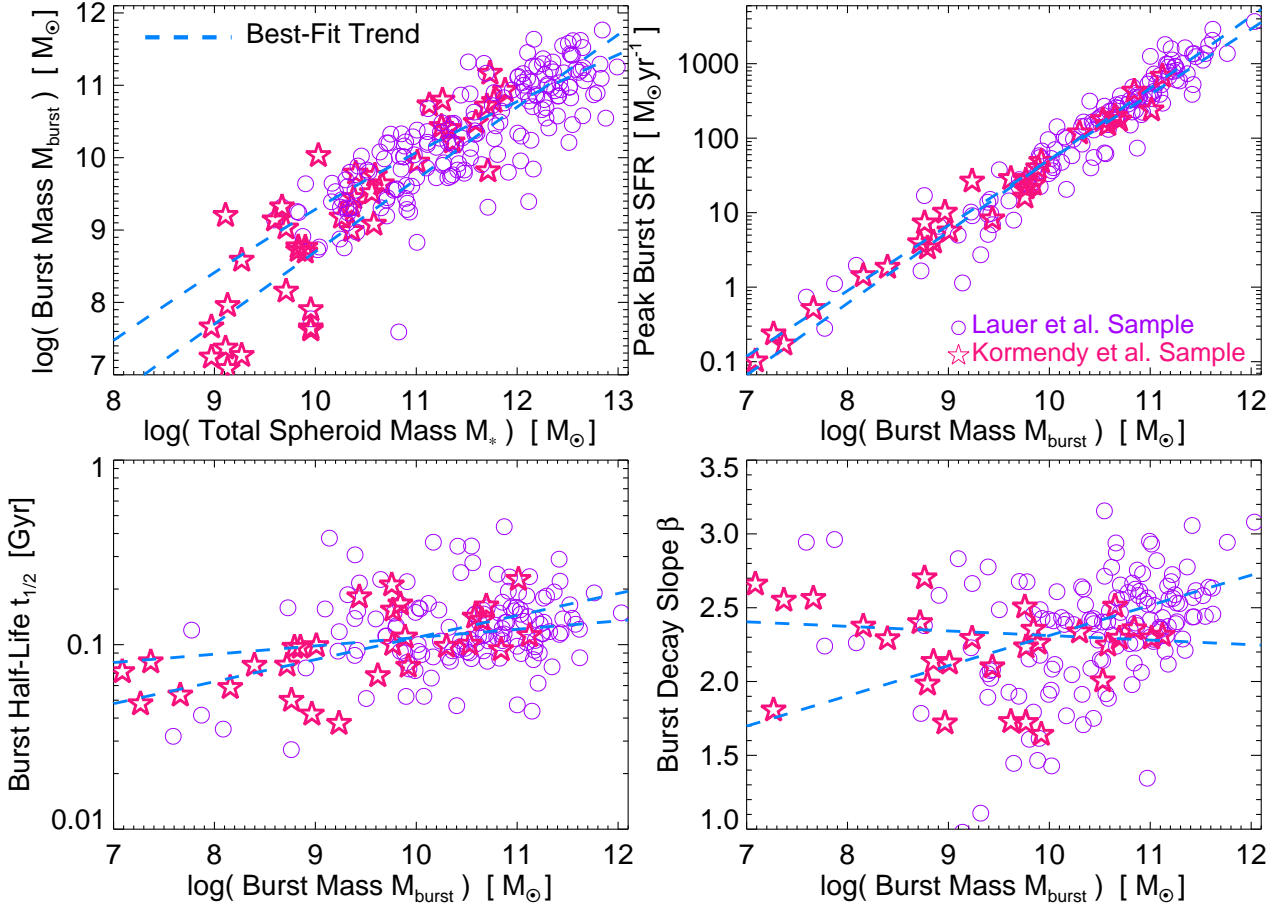}
    \caption{Observationally inferred properties of spheroid starbursts as a function of 
    mass, from the relic starbursts. 
    {\em Top Left:} Mass of the starburst component of spheroids
    as a function of 
    total spheroid stellar mass. Detailed determinations and tests of these component 
    decompositions are presented in H09b,e.
    We show results for  
    the Virgo galaxy sample of 
    \citet[][red stars]{jk:profiles} and local massive spheroid 
    sample of \citet[][violet circles]{lauer:bimodal.profiles}. Dashed blue lines show the 
    range of best-fit scalings to the observed systems (given in Table~\ref{tbl:fits}). 
    {\em Top Right:} Peak SFR of the starburst, as a function of total 
    starburst mass. 
    {\em Bottom Left:} Duration of the starburst (time from peak to half-peak SFR). 
    {\em Bottom Right:} Best-fit power-law slope to the decay of the SFR 
    versus time (Equation~\ref{eqn:mdot.time}). 
    \label{fig:fit.results}}
\end{figure*}

Figures~\ref{fig:fit.results} illustrates these 
fits and a number of global properties of 
the bursts, using our full sample of profiles taken from 
\citet{jk:profiles} and \citet{lauer:bimodal.profiles}. 
We also present the relevant fits and scaling relations 
in Table~\ref{tbl:fits}. 
First, Figure~\ref{fig:fit.results} shows the {\em total} 
burst mass (i.e.\ total mass of the inner/dissipational component) 
as a function of total galaxy stellar mass ({\em top left}). 
These values, and independent tests of their validity, as 
well as comparison to models of spheroid formation and 
observed disk gas fractions, are discussed 
extensively in \citet{hopkins:cusps.mergers,hopkins:cusps.ell,
hopkins:cores,hopkins:cusps.fp,hopkins:cusps.evol}. 
To lowest order, a fraction $\sim10\%$ is typically 
in the burst/dissipational component, i.e.\ 
$M_{\rm burst}\sim 0.1\,M_{\ast}$.  For somewhat higher 
accuracy, we use the fit from \citet{hopkins:cusps.fp}, 
motivated by comparison with disk gas fractions that 
constitute burst progenitors, 
\begin{equation}
M_{\rm burst} \approx \frac{1}{[1 + (M_{\ast}/10^{9.15}\,\msun)^{0.4}]}\ .
\label{eqn:mburst.vs.mass}
\end{equation}
This is also shown in Figure~\ref{fig:fit.results}, along with the 
simpler linear fit. 

\begin{footnotesize}
\ctable[
  caption={{\normalsize Starburst Scaling Relations}\label{tbl:fits}},center,star
  ]{llccc}{
\tnote[ ]{For the parameters $x$ and $y$, the best-fit linear scaling 
of $y(x)$ is presented, with the assumed functional form 
$y = a + b\,x$. Errors on the fitted parameters $a$ (normalization) 
and $b$ (slope) are given, along with the 
intrinsic scatter in $y$ about this correlation, $\sigma$.}
}{
\hline\hline
\multicolumn{1}{c}{$y$\tmark[\ ]} &
\multicolumn{1}{c}{$x$} &
\multicolumn{1}{c}{$a$} & 
\multicolumn{1}{c}{$b$} & 
\multicolumn{1}{c}{$\sigma$} \\
\hline
$\log{(M_{\rm burst}/10^{10}\,M_{\sun})}$ & $\log{(M_{\ast}/10^{11}\,M_{\sun})}$ 
& $-0.05\pm0.12$ & $0.78\pm0.16$ & $0.41$ \\ 
$\log{(\dot{M}_{\ast}[0]/M_{\sun}\,{\rm yr^{-1}})}$ & $\log{(M_{\rm burst}/10^{10}\,M_{\sun})}$ 
& $1.70\pm0.05$ & $0.92\pm0.04$ & $0.22$ \\ 
$\log{(t_{1/2}/{\rm Gyr})}$ & $\log{(M_{\rm burst}/10^{10}\,M_{\sun})}$ 
& $-0.96\pm0.04$ & $0.08\pm0.04$ & $0.20$ \\ 
$\log{(\dot{M}_{\ast}[0])}$ & $\log{(M_{\rm burst}/t_{1/2})}$ 
& $-0.24\pm0.01$ & $1.00\pm0.01$ & $0.04$ \\ 
$\beta$ & $\log{(M_{\rm burst}/10^{10}\,M_{\sun})}$ 
& $2.38\pm0.12$ & $0.09\pm0.11$ & $0.35$ \\ 
$\log{(R_{\rm relic}/{\rm kpc})}$ & $\log{(M_{\rm burst}/10^{10}\,M_{\sun})}$ 
& $-0.43\pm0.07$ & $0.58\pm0.07$ & $0.29$ \\ 
$\log{(R_{1/2}/{\rm kpc})}$ & $\log{(M_{\rm burst}/10^{10}\,M_{\sun})}$ 
& $-0.57\pm0.07$ & $0.60\pm0.06$ & $0.26$ \\ 
$\log{(t_{1/2,\,r}/{\rm Gyr})}$ & $\log{(M_{\rm burst}/10^{10}\,M_{\sun})}$ 
& $-1.07\pm0.08$ & $0.10\pm0.08$ & $0.28$ \\ 
$\beta_{r}$ & $\log{(M_{\rm burst}/10^{10}\,M_{\sun})}$ 
& $0.45\pm0.01$ & $-0.035\pm0.017$ & $0.11$ \\ 
$\log{({\rm Burst\ Age}/{\rm Gyr})}$ & $\log{(M_{\ast}/10^{11}\,M_{\sun})}$ 
& $0.95\pm0.11$ & $0.15\pm0.04$ & $0.12$ \\ 
${{\rm Burst\ Age}/{\rm Gyr}}$ & $\log{(M_{\ast}/10^{11}\,M_{\sun})}$ 
& $9.5\pm0.4$ & $3.0\pm0.6$ & $2.1$ \\ 
\hline\hline\\
}
\end{footnotesize}

Next, we show the burst half-life $t_{1/2}$ ({\em bottom left}), 
as a function of burst mass (the comparison versus total 
stellar mass is similar, but other possible correlations are less 
significant). 
To lowest order this is mass-independent, with a 
value $\sim10^{8}\,$yr -- depending on the definition, 
this equates to a total observable burst lifetime of a 
couple $10^{8}$\,yr with an inherent factor $\sim2-3$ scatter. 
Fitting to a power law, we find some (weak) mass dependence, 
\begin{equation}
t_{1/2} \approx 1.1\times10^{8}\,{\rm yr}\, 
{\Bigl (}\frac{M_{\rm burst}}{10^{10}\,\msun} {\Bigr )}^{0.1}\ ,
\label{eqn:t12.vs.mass}
\end{equation}
with a mass-independent scatter of 
$\sigma_{t_{1/2}} \approx 0.2$\,dex about this median relation.

This agrees well with expectations from e.g.\ numerical 
simulations and extrapolation of the 
\citet{kennicutt98} law, that the burst duration should reflect the dynamical 
times in the central regions of galaxies; 
as e.g.\ a disk galaxy dynamical time $\propto R_{e}/V_{c}$ 
is only weakly mass-dependent \citep[$\propto M_{\ast}^{(0.05-0.08)}$, 
if we compare $R_{e}\propto M_{\ast}^{0.31}$ and 
$V_{c}\propto M_{\ast}^{0.25}$ from][]{courteau:disk.scalings} 
\citep[see also][]{belldejong:tf,mcgaugh:tf,avilareese:baryonic.tf}.
Moreover, the scatter in 
the dynamical times expected from these correlations is similar 
to that in Figure~\ref{fig:fit.results}. 

We next show the slope of the power-law decay, 
$\beta$ ({\em bottom right}). There is no obvious correlation 
with burst mass or any other parameter; however, 
the values cluster in a reasonably narrow range, 
$\beta\sim2-3$, with a best-fit median of $\beta\approx2.38$. 
Again, this matches expectations; given some 
Schmidt-law star formation relation, the 
slope $\beta$ follows from the profile shape of the 
relic burst component. 
These profile shapes, and their physical origins, are discussed 
extensively in \citet{hopkins:cusps.mergers,hopkins:cusps.ell}; 
there, the authors show that they can be reasonably well-parameterized 
as Sersic functions with relatively low Sersic indices $n_{s}\sim1-3$, 
and that this follows naturally from the fact that they form 
via gas inflow and dissipation. Given such a Sersic-index profile, 
although there are no trivial analytic solutions for $\beta$, a simple 
numerical calculation shows that $\beta\approx2-3$ is the natural, 
corresponding expectation. 

We next consider the peak burst star formation 
rate, $\dot{M}_{\ast}(0)$ ({\em top right}). 
Immediately, it is clear that this is very tightly correlated with the 
total burst mass $M_{\rm burst}$. We compare the correlation 
expected from Equation~\ref{eqn:mdotmax}, assuming a 
typical $\beta\approx2.4$ such that 
$(\beta-1)\,(2^{1/\beta}-1)\approx0.5$, i.e.\ 
$\dot{M}_{\ast}(0) \approx 0.5\,M_{\rm burst}/t_{1/2}$. 
We show this both for an assumed constant 
$t_{1/2}=5\times10^{7}\,$yr, and for $t_{1/2}$ given 
by the weakly mass-dependent fit in Equation~\ref{eqn:t12.vs.mass}. 
Both cases agree very well with the observed trend, 
with a small, mass-independent scatter 
of $\sigma_{\dot{M}_{\ast}(0)}\approx0.15-0.2$\,dex. 

Worth noting is that, within massive observed systems, 
peak burst SFRs of $>1000\,\msun\,{\rm yr^{-1}}$ 
are obtained, but there is a large dynamic range -- low mass systems 
can have ``bursts'' with implied peak SFRs as low as 
$\sim0.1-1\,\msun\,{\rm yr^{-1}}$. 
Also, since there is relatively little variation in $t_{1/2}$ at fixed 
mass, the scatter is, as mentioned above, small. 
Producing a very high-SFR burst therefore requires an 
extremely high-gas-mass system; the case where a lower-mass system might 
compress its gas much more efficiently, leading to a shorter-lived but 
arbitrarily highly-peaked burst appears to be rare.

Finally, we note that their is some (factor $\sim2$) observational uncertainty in the exact 
normalization of the correlation between SFR and gas surface densities. 
Adopting a different IMF will likewise systematically shift the inferred relation 
from \citet{kennicutt98}.
However, so long as the change in Equation~\ref{eqn:kennicutt} is purely in 
normalization, it has no effect on the shape of the inferred lightcurves, 
and translates directly to corresponding normalization changes 
in the resulting SFR and timescales in Figures~\ref{fig:examples}-\ref{fig:fit.results.r} and 
Table~\ref{tbl:fits}. Specifically, if the normalization in 
Equation~\ref{eqn:kennicutt} is multiplied by a factor $\eta$, 
then the implied peak SFR $\dot{M}_{\ast}[0]$ increases by the same factor $\eta$, 
and the burst timescale $t_{1/2}$ (and $t_{1/2,\,r}$) 
decrease by a factor $\eta$. All other parameters 
in Table~\ref{tbl:fits} are unchanged.

\subsubsection{Spatial Sizes of Bursts}
\label{sec:results:properties:sizes}

\begin{figure*}
    \centering
    \plotside{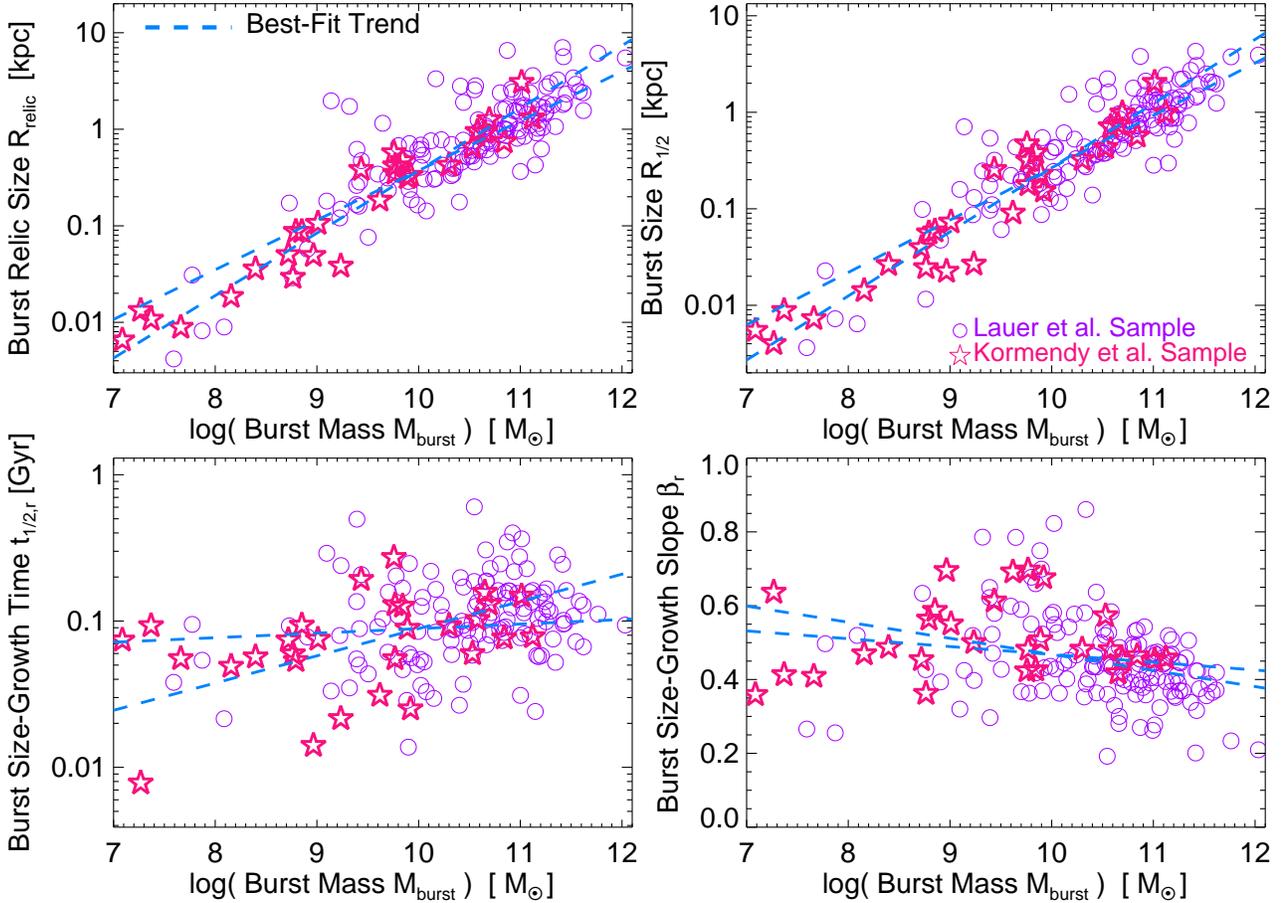}
    \caption{As Figure~\ref{fig:fit.results} -- observationally inferred properties of 
    spheroid starbursts as a function of 
    mass, from the relic starburst stars -- but here giving the spatial size distribution of starbursts. 
    {\em Top Left:} Size (projected half-stellar-mass radius) of the burst relics (generally a 
    factor $\sim0.1$ of the total galaxy size $R_{e}$), directly from the observed galaxy properties. 
    {\em Top Right:} Size of the starburst at time $t=t_{1/2}$ (when the SFR is half-peak). 
    Size here is defined as half-SFR or half-light (assuming light traces SFR) radius; hence is 
    more compact than the relic size. 
    {\em Bottom Left:} Size-doubling time of the starburst (time for e.g.\ relative gas exhaustion 
    in central, high density regions to leave only star formation at larger radii, giving a larger 
    apparent radius). This is simply related to the SFR decay timescale in Figure~\ref{fig:fit.results}. 
    {\em Bottom Right:} Best-fit power-law slope to the increase of starburst scale radius  
    versus time (Equation~\ref{eqn:r.time}). 
    \label{fig:fit.results.r}}
\end{figure*}

Figure~\ref{fig:fit.results.r} continues our analysis, plotting 
parameters as in Figure~\ref{fig:fit.results} 
but for the burst spatial sizes. 
First, we consider the effective (projected half-stellar mass) 
radius of the burst remnant ({\em top left}), 
i.e.\ the radius of {\em all} stars after the burst has completed, 
namely of the relic profile as shown in Figure~\ref{fig:examples} ({\em top left}). 
We consider this as a function of burst mass, and find a 
correlation of the form 
\begin{equation}
R_{\rm relic} \approx 0.37\,{\rm kpc}\,{\Bigl (} \frac{M_{\rm burst}}{10^{10}\,\msun} {\Bigr )}^{0.58}
\label{eqn:r.relic}
\end{equation}
with scatter $\sigma_{R_{\rm relic}}\approx0.3$\,dex.
As a property of the {\em total} 
mass distribution, it is studied in more detail in 
\citet{hopkins:cusps.mergers,hopkins:cusps.ell}. 
There, and in \citet{hopkins:ang.mom.overview}, it is shown that such sizes arise naturally 
from dissipational inflows, as a consequence of where the inflows 
stall and become self-gravitating. 
This is the reason that the power-law scaling in Equation~\ref{eqn:r.relic} 
reflects the observed scaling of spheroid sizes with mass 
\citep[see e.g.][]{shen:size.mass}, and
is a consequence of simple dynamics, independent of 
the star formation law. 

Next, we consider $R_{\rm burst}$ ({\em top right}), the effective 
(half-{\em light}, assuming light traces local SFR) radius of the burst itself 
during the starburst episode (as shown in Figure~\ref{fig:examples}; {\em bottom}). 
For convenience, and as a reference point, we define the plotted 
quantity as 
$R_{\rm burst}(t_{1/2})$, i.e.\ the size at time $t=t_{1/2}$. 
This is a convenient parameterization and also has the advantage of 
being more representative of what might be observed, since 
the duty cycle at peak is vanishingly small. 
In any case, there is a similar correlation, with 
\begin{equation}
R_{\rm 1/2}\equiv R_{\rm burst}(t_{1/2})
\approx 0.27\,{\rm kpc}\,{\Bigl (} \frac{M_{\rm burst}}{10^{10}\,\msun} {\Bigr )}^{0.60}
\label{eqn:r.burst}
\end{equation}
with $\sigma_{R_{\rm relic}}\approx0.3$\,dex scatter. 

As expected, the size $R_{\rm burst}$ near the peak of 
star formation 
is much smaller (by a typical factor $\sim$a few) than 
the size scale of the relic stars. This owes to two factors. 
First, near-peak, it is the highest-density material just at the 
galaxy center that dominates the SFR, while almost none 
of the more extended material is contributing at a significant level -- 
hence the size of the burst itself will still grow by a large factor with 
time (see Figure~\ref{fig:examples}). Second, 
this size is weighted by SFR, i.e.\ by gas density to 
some super-linear power according to the \citet{kennicutt98} 
relation. If $\Sigma_{\rm gas}$ follows a Sersic law 
with some index $n_{s}$ and effective radius $R_{e}$, 
and $\Sigma_{\rm SF}\propto \Sigma_{\rm gas}^{n_{K}}$,
then it is trivial to show that the half-SFR radius 
will simply be $R_{e}\times n_K^{-n_{s}}$. 
For $n_{s}=4$ and $n_{K}=1.5$, this gives a factor of almost 
$5$ smaller $R_{e}$ for the SFR distribution 
(for $n_{s}=2$, more typical of the inner components of the 
systems here, this is a more moderate factor $\sim2$). 

Nevertheless, the spatial sizes can still reach several kpc.
In particular, for the {\em same} systems that reach $>1000\,\msun\,{\rm yr^{-1}}$ 
peak SFRs, the effective radii are also large, $\sim1-10\,$kpc. 
Moreover, we refer here only to the strict half-SFR radius; 
estimating an observed size in detail requires 
radiative transfer modeling.  Analysis of simulations 
in \citet{wuyts:model.numbers.and.colors.vs.obs.prep,
wuyts:photometry.biases.mgrrem}, and similar analysis of some well-studied 
local starburst galaxies \citep{laine:toomre.sequence}, suggests that -- especially at the 
peak of activity -- the IR or radio size can be biased to factors of $\sim2$ 
larger than the true half-SFR radius owing to obscuration in the 
high-column-density nucleus and reprocessing of the emitted light. 

Next, we consider the burst size-doubling timescale, 
$t_{1/2,\,r}$. Unsurprisingly, the values are very similar to the SFR 
decay times $t_{1/2}$, with similar weak dependence on mass. 
We find a best-fit median $t_{1/2,\,r}$  
\begin{equation}
t_{1/2,\,r} \approx 8.5\times10^{7}\,{\rm yr}\, 
{\Bigl (}\frac{M_{\rm burst}}{10^{10}\,\msun} {\Bigr )}^{0.1}\ ,
\label{eqn:t12r.vs.mass}
\end{equation}
with a mass-independent scatter of 
$\sigma_{t_{1/2,\,r}} \approx 0.3$\,dex. 

We also consider the slope $\beta_{r}$, of the starburst size versus time. 
Here, there may be a weak correlation with mass at the high-mass 
end, but formally this is only marginally significant ($\sim2\,\sigma$) 
and only if we restrict ourselves
to $M_{\rm burst} \sim10^{10}-10^{12}\,\msun$. 
We therefore hesitate to quote a correlation; but do note the relatively 
robust median value of $\beta_{r}\sim0.4-0.5$, with 
scatter $\sigma_{\beta_{r}}\approx 0.1$. 
Again, these naturally follow from the combination of 
profile shapes and the \citet{kennicutt98} relation.

\subsubsection{Effects of a Steeper Kennicutt-Schmidt Relation}
\label{sec:results:properties:kennicutt}

Thus far, we have adopted the Kennicutt-Schmidt relation 
as calibrated at low redshifts (Equation~\ref{eqn:kennicutt}), 
with an index $n_{K}=1.4$ ($\dot{\Sigma}_{\ast}\propto \Sigma_{\rm gas}^{n_{K}}$). 
However, some recent observations of high-redshift, massively star-forming 
systems have suggested that the SFRs of such systems exceed the predictions 
with this index and favor a somewhat steeper slope 
of $n_{K}=1.7\pm0.1$ \citep{bouche:z2.kennicutt}. 
Specifically, they argue for a best-fit relation 
\begin{equation}
\dot{\Sigma}_{\ast} \approx 
9.3\times10^{-5}\,\msun\,{\rm yr^{-1}\,kpc^{-2}}
\,{\Bigl (}\frac{\Sigma_{\rm gas}}{\msun\,{\rm pc^{-2}}} {\Bigr )}^{1.7} \ .
\label{eqn:kennicutt.alt}
\end{equation}
We caution that the high-redshift observations -- 
in particular the inferences of both gas masses and 
total bolometric luminosities and hence SFRs -- remain uncertain. 
Nevertheless, this is an important source of error,
and generally larger than many of the other systematic uncertainties 
(those being at the factor of $\sim2$ level) in our modeling. 
Moreover, it can be particularly important in the most extreme 
systems, such as e.g.\ sub-millimeter galaxies, with high gas densities. 
We therefore re-consider our previous comparisons, 
but instead adopt this modified (steeper) index.

\begin{footnotesize}
\ctable[
  caption={{\normalsize Starburst Scaling Relations with 
  Alternative Kennicutt-Schmidt Relation}\label{tbl:fits.alt}},center,star
  ]{llccc}{
\tnote[ ]{As Table~\ref{tbl:fits}, but adopting 
a Kennicutt-Schmidt star formation versus gas surface density relation 
with a steeper index ($\dot{\Sigma}_{\ast} \propto \Sigma_{\rm gas}^{1.7}$), 
as suggested by observations of high-redshift, high-SFR systems.
}
}{
\hline\hline
\multicolumn{1}{c}{$y$\tmark[\ ]} &
\multicolumn{1}{c}{$x$} &
\multicolumn{1}{c}{$a$} & 
\multicolumn{1}{c}{$b$} & 
\multicolumn{1}{c}{$\sigma$} \\
\hline
$\log{(\dot{M}_{\ast}[0]/M_{\sun}\,{\rm yr^{-1}})}$ & $\log{(M_{\rm burst}/10^{10}\,M_{\sun})}$ 
& $2.62\pm0.12$ & $0.87\pm0.11$ & $0.35$ \\ 
$\log{(t_{1/2}/{\rm Gyr})}$ & $\log{(M_{\rm burst}/10^{10}\,M_{\sun})}$ 
& $-1.97\pm0.10$ & $0.11\pm0.09$ & $0.30$ \\ 
$\log{(\dot{M}_{\ast}[0])}$ & $\log{(M_{\rm burst}/t_{1/2})}$ 
& $-0.38\pm0.01$ & $1.00\pm0.01$ & $0.06$ \\ 
$\beta$ & $\log{(M_{\rm burst}/10^{10}\,M_{\sun})}$ 
& $1.69\pm0.05$ & $0.04\pm0.04$ & $0.22$ \\ 
$\log{(R_{1/2}/{\rm kpc})}$ & $\log{(M_{\rm burst}/10^{10}\,M_{\sun})}$ 
& $-0.67\pm0.07$ & $0.61\pm0.06$ & $0.26$ \\ 
$\log{(t_{1/2,\,r}/{\rm Gyr})}$ & $\log{(M_{\rm burst}/10^{10}\,M_{\sun})}$ 
& $-2.53\pm0.23$ & $0.06\pm0.21$ & $0.41$ \\ 
$\beta_{r}$ & $\log{(M_{\rm burst}/10^{10}\,M_{\sun})}$ 
& $0.33\pm0.01$ & $-0.02\pm0.01$ & $0.08$ \\ 
\hline\hline\\
}
\end{footnotesize}

\begin{figure}
    \centering
    \plotone{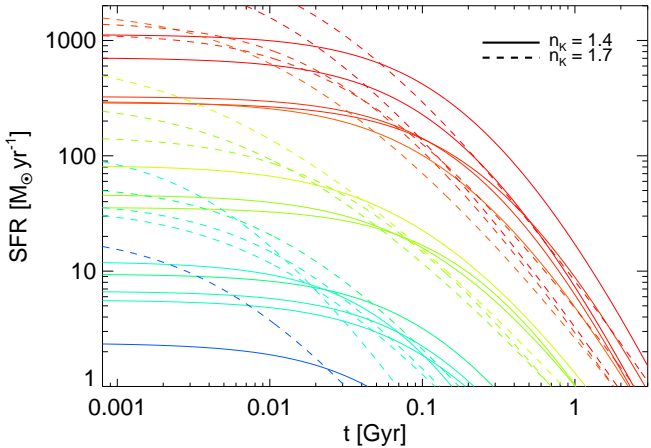}
    \caption{As Figure~\ref{fig:examples} ({\em middle}), 
    but comparing a subset of the inferred star formation histories from 
    the observed systems with different assumed Kennicutt-Schmidt law 
    indices. Solid lines, as Figure~\ref{fig:examples}, assume 
    the $z=0$ index $n_{K}\approx1.4$. Dashed lines assume a 
    steeper relation suggested by some high-redshift observations, 
    $n_{K}\approx1.7$. The latter leads to more sharply peaked bursts 
    with larger peak SFRs in early ($t <10^{8}\,$yr) stages. 
    The star formation histories after $\sim0.1\,$Gyr are similar, 
    once gas surface densities have decayed to more moderate levels. 
    Table~\ref{tbl:fits.alt} shows the results from re-fitting the correlations 
    in Figures~\ref{fig:fit.results}-\ref{fig:fit.results.r} with this higher 
    $n_{K}=1.7$. 
    \label{fig:highks.examples}}
\end{figure}

Table~\ref{tbl:fits.alt} shows the resulting revised fits, for quantities 
that are affected by this (e.g.\ the SFR and size versus time). 
The results are, in all cases, qualitatively very similar to those 
in Figures~\ref{fig:examples}-\ref{fig:fit.results.r}.   
We therefore show only the results of fitting the correlations, and 
present a simple 
illustration of how the inferred burst properties change with 
$n_{K}$ in Figure~\ref{fig:highks.examples}. Specifically, we 
repeat our analysis of the observed Virgo systems from Figure~\ref{fig:examples}, 
but include both the previous results (with $n_{K}=1.4$) 
and the steep-index results (with $n_{K}=1.7$). For clarity, we show only a 
random subset of the observed systems. 
There are significant, systematic offsets between the implied star 
formation histories. 

As expected, with a steeper index -- i.e.\ larger SFR at high 
densities, the {\em peak} SFR of bursts is systematically 
higher. The difference is large -- a factor of $\approx 5$ higher peak SFR 
at fixed burst mass -- giving 
$\langle \dot{M}_{\ast}(0)\rangle \approx 
420\,\msun\,{\rm yr^{-1}}\,(M_{\rm burst}/10^{10}\,\msun)^{0.87}$; 
in other words, this implies that 
most systems with $M_{\rm burst}>3\times10^{10}\,\msun$ 
will exceed $1000\,\msun\,{\rm yr^{-1}}$ SFRs at the peak 
of their starburst phase. 
However, with a higher SFR comes  
faster gas exhaustion, so the half-life of this peak, 
$t_{1/2}$, is systematically shorter, by a factor of $\approx 6$ -- 
i.e.\ typical exhaustion times $t_{1/2}\approx10^{7}\,$yr. 
The decay slope $\beta$ is also systematically different, 
now $\beta \approx 1.7$ (as opposed to $2.4$ before) -- 
this shallower power-law decay may appear counter-intuitive, but,
from the form of Equation~\ref{eqn:mdot.time}, relates to the 
fact that the SFR at $t \ll t_{1/2}$ is more peaked 
with $n_{K}=1.7$, and then at $t\gg t_{1/2}$ (when the 
gas density is lower) more extended (because the 
steeper index yields lower SFRs at low surface densities). 

In practice, if one defines a starburst ``duration'' by time spent 
above some fixed, high SFR, this change in the Kennicutt-Schmidt slope 
does not actually have as dramatic an effect as implied by the 
change in $t_{1/2}$. 
At $t>0.1\,$Gyr, in fact, there is relatively little difference in 
the SFR versus time, starting from the same observed 
relic starburst density profile (i.e.\ for the same relic object). 
Most of the difference comes at $t<0.1\,$Gyr. 
This is clear in Figure~\ref{fig:highks.examples}.

\subsection{Inferring the Burst Star Formation History}
\label{sec:results:history}

\subsubsection{Methodology: The Ages of Bursts Today}
\label{sec:results:history:methods}

Having quantified how burst properties scale with mass, 
and how these are connected to overall properties of their 
host galaxies, we require only a simple extension of our 
analysis to infer the cosmic history of starbursts. 

\begin{figure}
    \centering
    \plotone{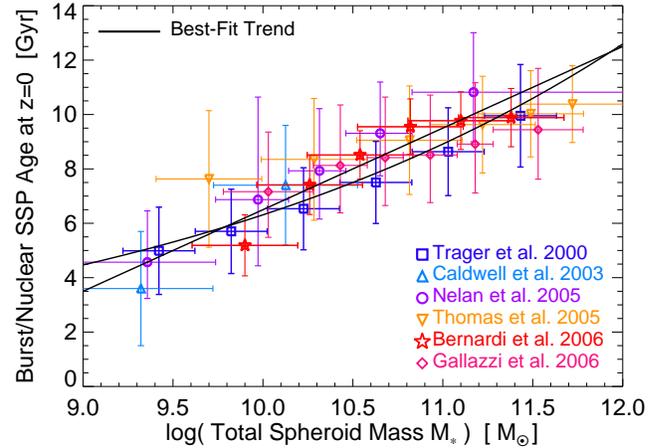}
    \caption{Observed average single stellar population age (at $z=0$) of 
    spheroid nuclear/most recent burst populations, as a function of 
    total spheroid stellar mass. Points show the median 
    age at each mass from various determinations; error bars show the 
    scatter (not the much smaller error in the mean). 
    Observations are compiled from various sources (labeled). 
    Solid black lines show the best-fit log-log and log-linear trends 
    for the median values
    (Equation~\ref{eqn:tburst.fit} and Table~\ref{tbl:fits}). 
    \label{fig:ages}}
\end{figure}

Figure~\ref{fig:ages} shows the observed ages, at $z=0$, of the 
dominant starburst in $z=0$ ellipticals/spheroids as a function of their 
total stellar mass. 
We compile observations from a number of sources 
\citep{trager:ages,caldwell:ssps,nelan05:ages,thomas05:ages,
gallazzi06:ages,bernardi:2006.ell.ages}; 
these involve various observational methodologies 
and selection functions, and together should plausibly represent the 
systematic uncertainties in this process. 
Note that for some of these samples, ages are quantified by 
environment; since the mass density of spheroids at the masses plotted 
is dominated by those in field environments, we show these values. 
Other observational estimates are within the scatter defined 
by these observations \citep[e.g.][]{jorgensen:1999.ell.ages.from.fp.evol,
kuntschner:ell.ages,gallazzi:ssps}. 
For each sample, the median age is well-determined and 
agrees reasonably well with other estimates.
A good approximation to the median ages is given by 
\begin{equation}
t_{\rm lookback,\,burst} \approx 8.9\,{\rm Gyr}\,
{\Bigl (}\frac{M_{\ast}}{10^{11}\,\msun}{\Bigr)}^{0.15}\ .
\label{eqn:tburst.fit}
\end{equation}
We also plot the $1\,\sigma$ scatter in ages at each mass determined 
from each sample; these too agree well. We 
find that this is reasonably parameterized as a mass-independent, 
Gaussian $\pm2\,$Gyr scatter. 

Various analyses, for example observationally analyzing mock 
SEDs constructed from hydrodynamic galaxy formation simulations 
or semi-analytic models of galaxy formation 
\citep[see e.g.][and references therein]{wuyts:model.numbers.and.colors.vs.obs.prep,
wuyts:photometry.biases.mgrrem,trager:test.ssp.fits.vs.sams}
have shown that, for typical spheroids that combine stars formed over 
an extended (potentially still short in absolute terms) 
star formation history in pre-merger disks (assembled dissipationlessly in mergers) 
with stars formed in bursts via gas dissipation in mergers, 
these observed ages do reflect the cosmic time at which those 
bursts occurred. Since a number of constraints and theoretical 
models show that the time of last major gas-rich merger 
and time when star formation in ``quenched'' systems 
must shut down are tightly coupled (even though there may not 
necessarily be a causal relationship between the two), this 
can also be thought of as dating the time of the
last burst of star formation 
that consumed the residual gas in the systems 
\citep[see][]{bundy:mtrans,bundy:agn.lf.to.mf.evol,
haiman:qlf.from.ell.ages,
hopkins:transition.mass,hopkins:red.galaxies,
cattaneo:sam,hopkins:groups.ell,
shankar:implied.msig.from.gal.ages}. 
Moreover, most of these ages are derived specifically from the {\em
central} stellar populations, those that are clearly formed
dissipationally; thus they represent a clear constraint on when this
burst occurred independent of more complex assembly and star formation
histories for the more extended components. Finally, where available,
detailed, resolved measurements
of stellar population gradients show
that these central populations do represent a
distinct component, in agreement with dissipational models, and
that their properties favor a single 
burst, rather than an extended star formation history or
dissipationless assembly of many distinct sub-components
\citep{schweizer96,titus:ssp.decomp,schweizer:7252,
schweizer:ngc34.disk,reichardt:ssp.decomp,michard:ssp.decomp}.

Given these comparisons, it seems reasonable to take the age
distributions in Figure~\ref{fig:ages} as a first approximation to the
times when these starbursts occurred.  Together with the analytic fits
to the distributions of burst properties versus mass, this enables us
to reconstruct the burst history of the Universe. Specifically, we
begin with the mass function of bulges/spheroids at $z=0$, here
adopting the determination in \citet{bell:mfs}. We ignore
disks/late-type galaxies and very low-mass bulges, as these contribute
negligibly to the quantities of interest (they have very low burst
masses, hence low peak SFRs, where we will show the population is
dominated by non-burst star formation).\footnote{
We have experimented with alternative determinations of the bulge-dominated galaxy 
mass function at $z=0$, including those presented in 
\citet{bernardi:gal.corr.by.type.and.mf}
and \citet{vika:2009.smbh.mf.from.lbulge}. 
We have also attempted to construct the mass function of 
bulges specifically, following \citet{driver:bulge.mfs} or 
by adopting the morphology-separated mass functions from 
\citet{kochanek:morph.stellar.mf} with a type-dependent $B/T$ from \citet{balcells:bulge.scaling}. 
We find these make little difference to any of our conclusions, as the 
resulting differences lie primarily in identification of low mass bulges 
or bulges in late-type galaxies which are negligible in the total IR luminosity 
density. 
}
At a given mass, we know the
number density of systems today, and the distribution of burst masses
$M_{\rm burst}(M_{\ast,\,\rm tot})$ (from e.g.\
Equation~\ref{eqn:mburst.vs.mass}, with appropriate scatter); i.e.\ we
know the (co-moving) total number density of bursts with various
masses that must have occurred before $z=0$ to account for those
systems.  We also know the distribution of cosmic look-back times when
those bursts happened, from Equation~\ref{eqn:tburst.fit} above
(again, with appropriate scatter).  This allows us to construct a
Monte-Carlo population of bursts, each with some total mass and taking
place (defined here as peaking) at some cosmic lookback time. Note
that this is independent of the state of the remainder of the galaxy at
this time -- so is not affected if e.g.\ the outer portions of
ellipticals are assembled much later \citep[see
e.g.][]{naab:size.evol.from.minor.mergers,
bezanson:massive.gal.cores.evol,
hopkins:cusps.evol,hopkins:maximum.surface.densities,hopkins:r.z.evol}.

We can then assign a full lightcurve to each burst, using the 
relations in \S~\ref{sec:results:properties} and Table~\ref{tbl:fits}; 
namely, knowing the burst mass, a 
duration $t_{1/2}$ (Equation~\ref{eqn:t12.vs.mass}) and 
power law decay slope (median $\approx2.4$; again with scatter). 
Together these define the corresponding peak SFR $\dot{M}_{\ast}(0)$. 
The time $t=0$ in the lightcurve is simply the time of the burst, 
as determined above, and it evolves forward from this time 
(the exact choice makes little difference, since the burst durations
are small compared to the Hubble time). 
For convenience, we truncate each burst at a time 
$\approx 20\,t_{1/2}$, but by this point the luminosity has decayed sufficiently 
that the burst is negligible. 
We then make mock observation of the Monte Carlo population 
at any redshift $z$, and construct the luminosity function or luminosity 
density of bursts. Of course, we are really predicting a distribution of 
star formation rates; for comparison with observations, we convert these 
to total ($8-1000\,\mu{\rm m}$) infrared luminosities with the 
empirically-calibrated conversion factor in Equation~\ref{eqn:mdot.to.lir}. 
\footnote{Note that, in comparing to 
observed starburst light curves or luminosity functions, the 
average $M_{\ast}/L$ will depend systematically on the IMF. However, 
the inferred stellar surface mass density on which we base our 
analysis is determined from observed luminosity profiles, 
with the same systematic dependence on $M_{\ast}/L$. Thus, so 
long as the IMF does not evolve with redshift, variations between 
typical choices make little difference.}

We can also trivially repeat this procedure for a different bolometric
correction from $\dot{M}_{\ast}$ to $L_{\rm IR}$ (which will
systematically shift the predicted luminosities by some factor
$\sim2$).  Likewise, we can repeat our calculation using the fitted
scaling relations appropriate for a steeper Kennicutt-Schmidt law with
index $n_{K}\approx1.7$ rather than $n_{K}\approx1.4$, given in
Table~\ref{tbl:fits.alt}.

\subsubsection{The Luminosity Function of Bursts}
\label{sec:results:history:lfs}

\begin{figure*}
    \centering
    \plotside{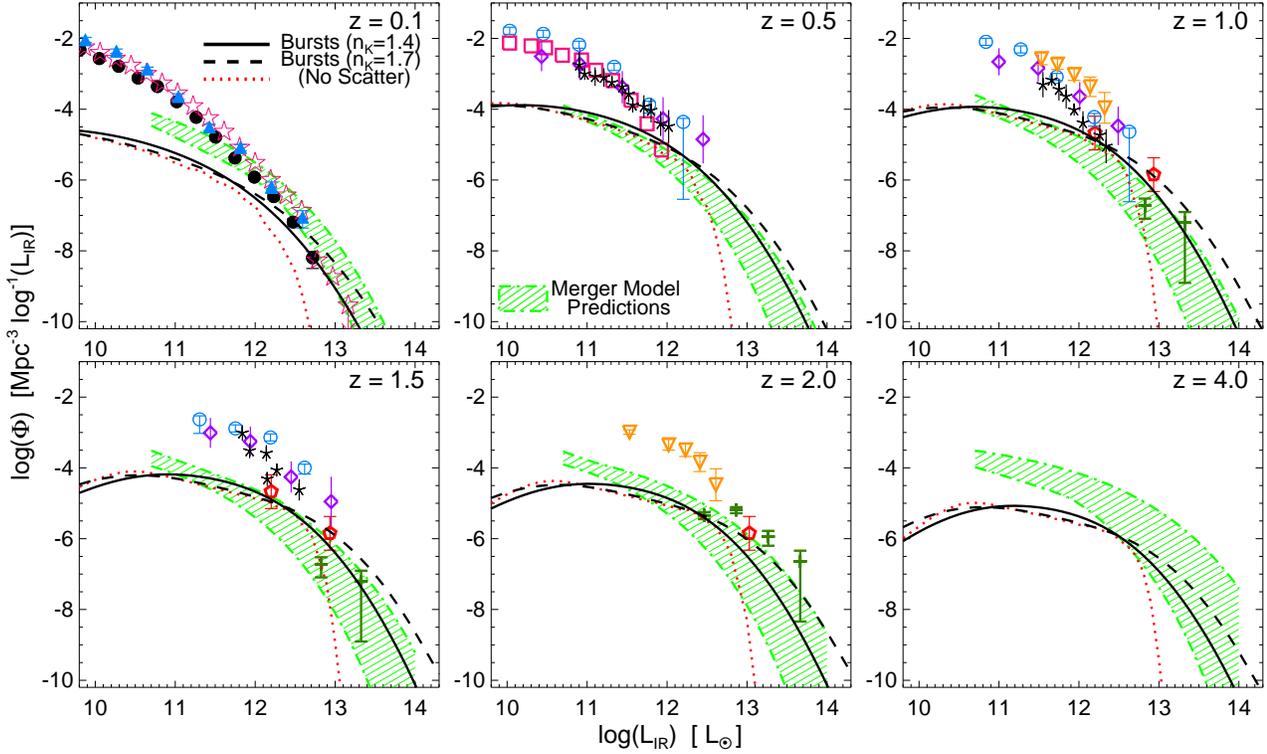}
    \caption{Total ($8-1000\,\mu{\rm m}$) IR luminosity functions as a function of redshift. 
    We show the contribution from bursts, using the methodology presented 
    here (black lines), assuming a Kennicutt-Schmidt index of $n_{K}=1.4$ (solid) 
    or $1.7$ (dashed). Dotted red line shows the result neglecting the scatter 
    in the correlations shown in Figures~\ref{fig:fit.results}-\ref{fig:fit.results.r}; 
    including the observed dispersion is important to explain rare, extreme starbursts. 
    Points show observed estimates from 
    \citet[][magenta stars]{saunders:ir.lfs}, 
    \citet[][blue triangles]{soifer:60m.lfs}, 
    \citet[][black filled circles]{yun:60m.lfs}, 
    \citet[][magenta squares]{huang:2007.local.ir.lf}, 
    \citet[][violet diamonds]{lefloch:ir.lfs},
    \citet[][orange inverted triangles]{caputi:ir.lfs}.
    \citet[][black $\ast$'s]{magnelli:z1.ir.lfs}
    \citet[][red pentagons]{babbedge:swire.lfs}, 
    \citet[][dark green $+$'s]{chapman:submm.lfs}, 
    and 
    \citet[][blue open circles]{perezgonzalez:ir.lfs}.
    Shaded green range shows the prediction (with systematic uncertainty) for 
    the IR LF of merger-induced bursts from the semi-empirical models and 
    hydrodynamic simulations in 
    \citet{hopkins:ir.lfs}$^{\ref{foot:mergercalc}}$.  
    The burst contribution dominates the bright end of the IR LF, 
    agreeing well with predicted merger-induced bursts and other constraints, 
    but is a small fraction of the typical $\sim L_{\ast}$ activity at any redshift. 
    The luminosity threshold above which bursts are important increases with 
    redshift, along with the entire LF. Fits to the burst LF 
    inferred here are provided in Table~\ref{tbl:fit.lfs} 
    \label{fig:ir.lfs}}
\end{figure*}

Figure~\ref{fig:ir.lfs} shows the resulting burst luminosity functions
at a variety of redshifts from $z=0-4$.  We compare with observations
compiled from a number of sources, available from $z\sim0-3$.  Note
that all of these are corrected to a total IR luminosity from
observations in some band; we adopt the corrections compiled in
\citet{valiante:ir.lfs.and.numbercounts}, but emphasize that some
caution, and at least a systematic factor $\sim2$ uncertainty in
$L_{\rm IR}$, should be considered in estimates from most if not all
observed wavelengths.

At low luminosities, the predicted LFs are well below those observed,
but they grow rapidly in importance, and are roughly consistent with
the observations, at the high-luminosity end.  This is expected -- it
is well-established that star formation at relatively low rates is
dominated by quiescent star formation in normal (e.g.\ non-merging)
galaxies; i.e.\ distributed star formation in disks, rather than
dissipational starbursts \citep[see e.g.][]{sanders96:ulirgs.mergers,
tacconi:smg.mgr.lifetime.to.quiescent,noeske:2007.sfh.part1,
noeske:sfh,bell:morphology.vs.sfr,jogee:merger.density.08,
robaina:2009.sf.fraction.from.mergers,
veilleux:ulirg.to.qso.sample.big.mdot.changes,sajina:pah.qso.vs.sf}.
At low redshifts, for example, nuclear starbursts (typically
merger-induced) are negligible at luminosities $\lesssim
10^{11}\,\lsun$; but at the highest luminosities, $>10^{12}\,\lsun$,
they become dominant \citep{sanders96:ulirgs.mergers}.  At higher
redshifts, the luminosity function of bursts increases rapidly, as
does the global luminosity function. This again is expected, as
increasing gas fractions and specific star formation rates lead all
systems (mergers and quiescent disks) to higher SFRs at fixed mass.
Thus, at {\em all} redshifts the (relatively) low-luminosity
population remains non-burst dominated, and the threshold (in terms of
$L_{\rm IR}$) for burst domination moves up to higher luminosities
($L_{\rm IR}\gtrsim 10^{12}\,\lsun$ at $z\sim1$, and $L_{\rm
IR}\gtrsim 10^{13}\,\lsun$ at $z\sim2$). In terms of space densities,
the burst population becomes important at $\sim 10^{-5}\,{\rm
Mpc^{-3}\,\log^{-1}{L_{\rm IR}}}$, with much weaker redshift
dependence.  This is similar to the behavior predicted in models of
merger-induced star formation bursts \citep{hopkins:ir.lfs}.

We show predictions for both the cases of the low-redshift
Kennicutt-Schmidt law slope of $n_{K}\approx1.4$, and for the
suggested steeper value of $n_{K}\approx1.7$ from high-redshift
observations.  Unsurprisingly, assuming a steeper relationship implies
more intense SFRs at the peak of activity, so the predicted burst LF
extends to higher luminosities.  However, the shorter gas exhaustion
time means that there is a lower number density at low/intermediate
luminosities.  The difference is clearly non-trivial, given the rarity
of the highest-luminosity sources. The results in the steep-index case
are comparable, however, to those obtained from the shallow-index case
if we were to assume the scatter in the relevant quantities is
systematically larger by $\sim0.1\,$dex -- in other words, this
continues the same trends, and the effects may be comparable in
magnitude to other uncertainties discussed here.

We also provide simple fits to the inferred luminosity function of bursts, as a function 
of redshift, for both the shallow and steep Kennicutt-Schmidt slopes. 
Following standard convention, we fit the luminosity function with a double power 
law form, 
\begin{equation}
\Phi \equiv \frac{{\rm d}n}{{\rm d}\,\log{L}} = 
\frac{\phi_{\ast}}{(L/L_{\ast})^{\alpha} + (L/L_{\ast})^{\gamma}} 
\label{eqn:double.pwr.law}
\end{equation}
where the parameters $\phi_{\ast}$ (normalization), 
$L_{\ast}$ (break luminosity), $\alpha$ (faint-end slope, 
i.e.\ $\Phi\propto L^{-\alpha}$ for $L\ll L_{\ast}$), 
and $\gamma$ (bright-end slope, 
i.e.\ $\Phi\propto L^{-\gamma}$ for $L\gg L_{\ast}$) 
depend on redshift, with that dependence conveniently approximated as
\begin{align}
\nonumber \log{L_{\ast}} &= L_{0} + L^{\prime}\,\xi + L^{\prime\prime}\,\xi^{2} \\ 
\nonumber \log{\phi_{\ast}} &= \phi_{0} + \phi^{\prime}\,\xi + \phi^{\prime\prime}\,\xi^{2} \\ 
\nonumber \alpha &= \alpha_{0} + \alpha^{\prime}\,\xi + \alpha^{\prime\prime}\,\xi^{2} \\ 
\nonumber \gamma &= \gamma_{0} + \gamma^{\prime}\,\xi + \gamma^{\prime\prime}\,\xi^{2} \\ 
\xi &\equiv \log{(1+z)}
\label{eqn:param.z.evol}
\end{align}
(Note that log here and throughout refers to $\log_{10}$.)  We perform
this fit using our results only up to redshift $z=4$, as the
uncertainties grow rapidly at higher redshift.
The fits should be considered with caution at higher redshifts.  We
provide the best-fit parameters in Table~\ref{tbl:fit.lfs}, along with
their uncertainties.

We note briefly that, especially because bursts are important at the
high-luminosity end of the total LF, incorporating the scatter in the
relevant relationships in \S~\ref{sec:results:properties} is critical
to the predicted abundance of IR-bright systems.  In
Figure~\ref{fig:ir.lfs}, we compare the prediction if we were to
ignore all scatter -- i.e.\ simply construct our Monte Carlo
population using just the median values of all parameters and their
correlations. As expected, this cuts off much more quickly at the
high-luminosity end, reflecting the rapid exponential cutoff in the
observed number of high-mass galaxies.

Recently, \citet{hopkins:ir.lfs} presented predictions from galaxy
formation models for the contribution to the SFR and IR luminosity
distributions from normal, quiescent star-forming galaxies, from
merger-induced bursts of star formation, and from AGN. The models used
a halo-occupation based approach to populate galaxies at each redshift
(i.e.\ simply beginning with observed galaxy stellar mass functions
and gas masses, with merger rates determined from evolving this
forward in agreement with observed merger fractions), then mapping
each population to suites of high-resolution simulations, to predict
the distribution of SFRs in various systems.  We compare their
predictions for the merger-induced starburst population to our
inferred burst SFR and IR luminosity distributions in
Figure~\ref{fig:ir.lfs}.\footnote{\label{foot:mergercalc}The merger rates determined in this 
model are presented in \citet{hopkins:merger.rates}. 
A ``merger rate calculator'' script to give the merger rate as a function of 
galaxy mass, mass ratio, and gas fraction, which determines these 
luminosity functions, is publicly available at 
\mergercalcurl.}
Given the systematic uncertainties they
quote (shown as the shaded range in the figure), and ours here, the
agreement is reasonable. The two diverge at $z\gg3$, but this is where
the uncertainties in both the modeling and our empirical inferences
become large. The faint-end extrapolations are also different, but
in neither case are these well-constrained (in a systematic sense, at
very low SFR/late times, the designation as ``burst'' is somewhat
arbitrary). In general, the agreement seen supports our interpretation
of the burst components of galaxies, and favors a possible merger
origin for these bursts.  More important, the burst history inferred
here should {\em not} correspond to the normal or quiescent star
forming population.

\subsubsection{The Problem of Bright SMGs}
\label{sec:results:history:smgs}

We also note that, at $z\sim2-3$, our reconstruction appears to
somewhat under-predict the abundance of the most luminous starburst
systems relative to observations.  However, at this redshift, the only
constraints at the high luminosities of interest come from the
sub-millimeter populations observed in \citet{chapman:submm.lfs}.
This is a relatively small sample, with a number of difficult
completeness corrections involved in estimating the number density,
and non-trivial cosmic variance given the sample selection.  Recently,
analyses of the number counts of similarly bright sources in much
larger IR surveys have suggested that the average counts may actually
be much lower -- a factor $\sim5$ difference \citep[with higher
estimates owing to substantial cosmic variance; see
e.g.][]{austermann:2009.aztec.submm.source.counts}.

Perhaps more important, given that the number densities in this regime
fall rapidly with $L_{\rm IR}$, a small error in the bolometric
correction translates to order-of-magnitude differences in number
density at fixed $L$. And indeed, such a conversion from the
sub-millimeter to total IR flux is highly uncertain, depending on
quantities that are not well-known for these populations, such as the
dust temperature, and subject to large object-to-object variation in
the local Universe.  
It should also be stressed that the observed high-redshift points 
in Figure~\ref{fig:ir.lfs} have not been corrected for possible AGN 
contributions to the IR luminosity, although (from indirect constraints) 
it has been argued that AGN are unlikely to contribute more 
than $\sim30\%$ of the bolometric luminosity in these systems 
\citep{menendezcelmestre:extended.submm.galaxies,casey:highz.ulirg.pops,
bussmann:2009.dog.luminosities}. 
For these reasons, we consider the comparison at
these redshifts and luminosities to be largely qualitative.  A more
detailed comparison would require full modeling from e.g.\
high-resolution simulations that include spatially dependent,
multi-phase star formation, metallicity, gas, and dust distributions
that can forward-model the full SED of such bursts \citep[for some
preliminary such results, we refer to][]{li:radiative.transfer,
jonsson:sunrise.release,narayanan:smg.modeling,
narayanan:molecular.gas.in.smgs,younger:warm.ulirg.evol}.

But the comparison in Figure~\ref{fig:ir.lfs} emphasizes an important
point: {\em if} the number density and bolometric corrections of such
high-luminosity systems are correct, then either the
\citet{kennicutt98} law must break down severely at high redshifts
\citep[giving much higher star formation rates than implied by e.g.\
observations of high-density systems in][at fixed surface
density]{bouche:z2.kennicutt}, or something must be fundamentally
wrong in our conversion between mass and light, either at low
redshifts (i.e.\ some dramatic errors in measurements of the surface
brightness and stellar masses of local ellipticals, which appears
unlikely) or at high redshifts (i.e.\ the results of a strongly
time-dependent stellar IMF, or large heavily-obscured AGN
contributions to the high-$z$ luminosities).

Taken at face value, the number density of high-$z$ bursts inferred,
coupled with their implied star formation rates, would imply far too
much high-density material in massive systems today.  All of our
analyses of burst star formation rates effectively set an upper limit
on the burst number density at high-$L$.  Consider, for example, the
consequences of relaxing our assumptions.  If starbursts were split
into several separate events, with lower initial burst rates, then the
duty cycle would go up, but the absolute $L$ would go down (and in
this portion of the luminosity function, the net result would be a
severe decrease in the high-$L_{\rm IR}$ prediction).  If systems were
not initially gas-dominated or $100\%$ gas, but maintained some lower
gas fraction in quasi steady-state for the time needed to build up
their densities, then again the peak luminosities decrease
severely. If the present-day densities are the result of assembling
many systems, this is the same as breaking up bursts, and leads to a
lower prediction. The bursts cannot be more concentrated in time
(yielding higher peak SFR) without violating the \citet{kennicutt98}
law; even doing so, they become more luminous but shorter, and it
requires an order-of-magnitude change to the star formation relation
before this yields a match.

Ultimately, a number of other obvious comparisons make this clear.
Yielding the implied SFRs of the observed bright SMG systems (taken at
face value) without breaking the \citet{kennicutt98} relation requires
$\sim 10^{11}\,\msun\,{\rm kpc^{-2}}$ gas (and ultimately stellar)
surface densities, over spatial radii of $\sim1-2\,$kpc.  Although
this is indeed comparable to the highest stellar surface densities in
ellipticals, observed systems only reach those surface densities at
$\lesssim100\,$pc scales; {\em not} at kpc scales (where the typical
surface density in a massive elliptical is $\approx
10^{10}\,\msun\,{\rm kpc^{-2}}$).  Also, given the characteristic SMG
lifetime of $\sim10^{8}\,$yr, from observational constraints, it is
straightforward to estimate the total amount of stellar mass that
would be formed at or above these surface densities, from SMGs over
the redshift range $z\sim2-4$, and the number comes to $\sim
10^{8}\,\msun\,{\rm Mpc^{-3}}\, [\Sigma_{\ast}>10^{11}\,\msun\,{\rm
kpc^{-2}}]$.  \citet{hopkins:density.galcores} calculate the
actual observed stellar mass density in all galaxies above such a
threshold, and find that it is $\approx 1.5\times10^{6}\,\msun\,{\rm
Mpc^{-3}}$, well below the range of uncertainties owing to e.g.\
stellar mass loss and the somewhat uncertain values above.

\subsubsection{The Burst Luminosity Density}
\label{sec:results:history:density}

\begin{figure*}
    \centering
    \plotside{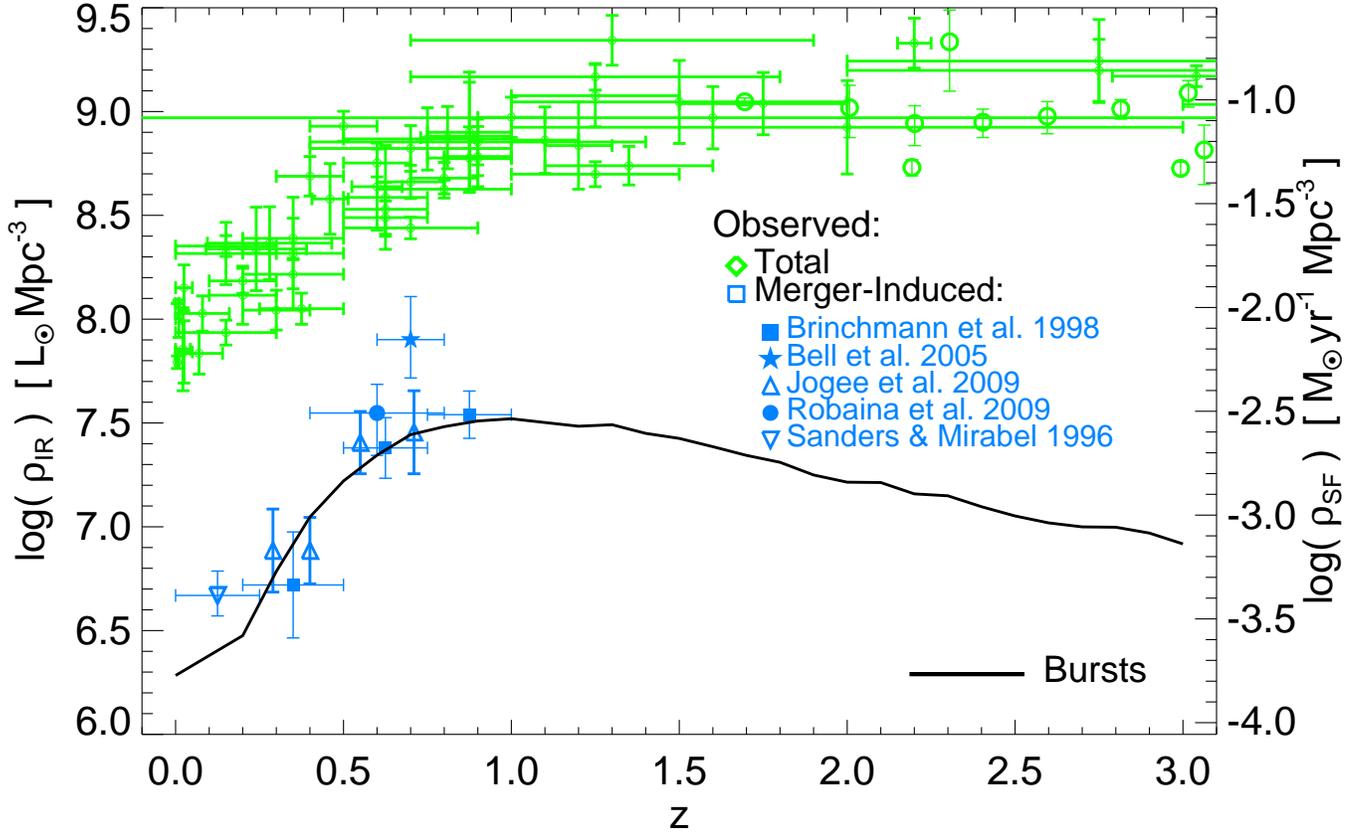}
    \caption{Total IR luminosity density (and corresponding SFR density) as a function of redshift. 
    We show the inferred contribution owing to bursts (solid black line) from the 
    observed $z=0$ spheroid properties. 
    We compare the observational compilation from \citet{hopkinsbeacom:sfh} 
    for the {\em total} IR luminosity density/SFR density (green diamonds), 
    and at high redshifts the SFR density inferred from Lyman-$\alpha$ forest 
    measurements in \citet[][green circles]{faucher:ion.background.evol}. 
    We also compare observed estimates of the SFR density 
    induced in mergers (blue points), from
    \citet[][triangles]{jogee:merger.density.08},  
    \citet[][inverted triangle]{sanders96:ulirgs.mergers}, 
    \citet[][circle]{robaina:2009.sf.fraction.from.mergers}, 
    \citet[][squares]{brinchmann:1998.morph.contrib.to.sfr}, 
    and \citet[][star]{bell:morphology.vs.sfr}.
    The burst component of star formation is a small fraction 
    $\sim5-10\%$ of the total IR luminosity density (most stars are formed 
    quiescently in disks and are then violently relaxed into spheroids). This 
    agrees well with observational estimates of the fraction of star formation 
    owing specifically to mergers, as predicted if these mergers induce the 
    angular momentum loss that drives gas dissipation in forming spheroids.  
    \label{fig:ir.lumden}}
\end{figure*}

Figure~\ref{fig:ir.lumden} integrates the luminosity functions shown in 
Figure~\ref{fig:ir.lfs} to give the total IR luminosity density, 
and corresponding total SFR density, owing to bursts, as a function of 
redshift. Note that many of the uncertainties that might affect our other results are 
integrated out here (e.g.\ the exact lightcurve shape, or 
Kennicutt-Schmidt slope). 
We compare with the total SFR density from a number of observations 
in IR, radio, and UV wavelengths
as compiled in \citet{hopkinsbeacom:sfh}, 
as well as that inferred at high redshifts from observations of the Lyman-$\alpha$ forest 
in quasar spectra and implied ionizing background, compiled 
in \citet{faucher:ion.background.evol}. 
As expected from Figure~\ref{fig:ir.lfs}, the predicted SFR density owing to 
bursts is well below the total SFR density observed. 
This clearly demonstrates that bursts do not, and cannot, dominate the 
total SFR density. 

We compare with a number of observations that attempt to estimate the
luminosity or SFR density induced by mergers.  First, we stress that
this is different from the {\em total} SFR density in objects {\em
identified as} mergers.  A merger can take $\sim2$\,Gyr to complete,
but the SFR over most of that time will simply reflect that of the
quiescent disks, with at most a modest enhancement. The burst itself,
where gas dissipation drives inflows into the center of the merger
remnant, enhancing the SFR significantly, has a duration of only
$\sim0.1\,$Gyr.  Nevertheless, using statistical samples of mergers
and non-merging systems, and comparing their SFR distributions,
various authors have attempted such a comparison.\footnote{Note that
most of these authors actually measure the {\em fraction} of the SFR
density in or induced by mergers, not the absolute value. We convert
this to an absolute density by rescaling with the observed total SFR
density at the same redshift from the best-fit observed trend
presented in \citet{hopkinsbeacom:sfh}.}

Typically, in these cases, the SFR density of some merger sample
(identified in a similar manner) is considered, but only after
subtracting away the expected contribution from quiescent star
formation. In general, this is accomplished via comparison to some
control sample of star-forming galaxies with similar stellar masses
and at comparable redshifts.
\citet{robaina:2009.sf.fraction.from.mergers} attempt this from a
pair-selected sample at $z\sim0.4-0.8$;
\citet{jogee:merger.density.08} perform such an estimate from
morphologically selected galaxies at $z\sim0.4-1$.  We also compare
with somewhat less well-defined, but similar samples that estimate the
total amount of star formation in observationally identified ongoing
mergers or recent (morphologically disturbed) merger remnants. We
compile observations from \citet{brinchmann:1998.morph.contrib.to.sfr}
and \citet{bell:morphology.vs.sfr}, who estimate this quantity in
morphologically-selected objects at $z\sim0-1.5$. 
We perform a similar exercise at low redshift using the fraction of late-stage major merger systems as a 
function of IR luminosity from \citet{sanders96:ulirgs.mergers}, together 
with the IR luminosity functions from \citet{saunders:ir.lfs} and 
\citet{yun:60m.lfs}, to estimate the fraction of 
the luminosity density in mergers. Because the
selection of ongoing mergers is somewhat strict in these samples
(isolating near-peak times), they are not very different from the
estimates of merger-induced star formation, but formally should still be 
considered upper limits. 

Our prediction agrees well with these observations, over the range
$z=0-1$. In fact, given the (probable factor $\sim2$) systematic
uncertainties involved in both, the level of agreement is surprising,
and may be somewhat coincidental. But, in general, both direct
estimates and our indirect constraint imply that bursts constitute
$\sim5-10\%$ of the SFR density, and that this is not dramatically
larger at high redshifts.  The agreement here also provides further
evidence that most of the bright, massive bursts of interest here are
really merger-induced, since that is specifically what is observed.
If there were a large non-merger population, the observations should
be much lower than our prediction.

We have also compared both to the predictions of the most recent
generation of cosmological models. \citet{somerville:new.sam} use
semi-analytic models, and
\citet{hopkins:groups.qso,hopkins:disk.survival.cosmo} employ
semi-empirical (halo-occupation based models), both combined with the
predictions from high-resolution galaxy merger simulations
\citep{cox:massratio.starbursts,hopkins:disk.survival} to predict the
merger-induced burst SFR and luminosity density as a function of
redshift.  Most recently, \citet{hopkins:ir.lfs} present a revised
version of these semi-empirical models based on larger suites of
simulations and improved halo occupation constraints and modeling, and
specifically present fits to the merger-induced burst population.
These are compared to our predicted luminosity functions in
Figure~\ref{fig:ir.lfs}. Comparing all of these results to our
predictions in Figure~\ref{fig:ir.lumden}, we find good agreement 
(unsurprising, given that the predicted luminosity functions in 
Figure~\ref{fig:ir.lfs} also agree well). The important point is that 
most recent models also predict that only $\sim5-10\%$ of the SFR density 
comes from mergers, at all redshifts. 

Finally, we hesitate to extrapolate our results beyond $z\sim2-3$, for
two primary reasons. First, because the Hubble time is short at higher
redshifts, the predictions become sensitive to $\sim0.1-1\,$Gyr
systematic uncertainties in stellar population age estimates, which
are known to be problematic particularly in the old-age
regime. Second, the higher in redshift we consider, the more likely we
are to see effects of differences between star formation and assembly
histories, even for the bursts themselves (where, at least at low
redshifts, observations suggest the two are tightly coupled; i.e.\
bursts tend to be dominated by a single most massive burst, rather
than many smaller sub-components). 
We therefore caution against the over-interpretation or extrapolation 
of our results at $z\gtrsim4$.

\section{Discussion}
\label{sec:discussion}

It has long been believed that the centers of galaxy spheroids must be
formed in dissipational starburst events, such that gas in the outer
parts of a galaxy must lose angular momentum rapidly, and fall in on
roughly a dynamical time to the galaxy center. Recently, observations
have shown that it is possible to robustly separate the ``burst''
component of galaxy profiles from the outer, violently relaxed
component owing to the scattering of progenitor galaxy stars formed
over earlier, more extended periods.  As detailed, high-resolution
observations of e.g.\ spheroid stellar mass profiles, shapes,
kinematics, and stellar population properties improve, such
decompositions become increasingly robust and applicable to a wide
range of systems.

In this paper, we use these observations as a novel, independent
constraint on the nature of galactic starbursts. We stress that by
``starburst'' here, we do {\em not} mean simply any high-SFR system;
rather, we refer specifically to star formation in central, usually
sub-kpc scale bursts owing to large central gas concentrations driven
by angular momentum loss as described above.  These are commonly
associated with galaxy-galaxy mergers, which have long been known to
efficiently drive starbursts and violent relaxation
\citep[e.g.][]{lynden-bell67,
toomre77,barnes.hernquist.91,barneshernquist96,barnes:review,
mihos:starbursts.94,mihos:starbursts.96}. But they could also owe to
any other sufficiently violent process; e.g.\ gas inflows owing to
disk bars or during dissipational collapse.

Regardless of origin, we can robustly identify the starburst relics
and stellar mass surface density profiles in well-observed spheroids
at $z=0$.  This methodology and tests of its accuracy have been
discussed in
\citet{hopkins:cusps.mergers,hopkins:cusps.ell,hopkins:cores,hopkins:cusps.fp}.
This alone is a powerful integral constraint on the star formation
histories of these systems. But we show that they can in fact be used
to provide more detailed information, by allowing for the inversion
and recovery of the star formation history of each burst. We can thus
use local observations to recover the full time-dependent and
scale-dependent star formation history of each burst, if we
couple these observed constraints to simple, well-motivated
assumptions.  Namely, that some form of the Kennicutt-Schmidt relation
(between star formation rate and gas surface density) applied in the
formation of these stars, and that they formed in dissipational (i.e.\
rapid and initially gas-rich, on these scales) events. Both
assumptions are directly motivated by observations, but we also
consider how uncertainties in them translate into uncertainties in the
resulting constraints.

Performing this exercise, we recover a large number of empirically
determined parameters of starbursts, and quantify how they scale as a
function of mass and other properties.  This includes the total mass
of the starburst and the time-dependent SFR. In general, we find that
the constrained starbursts can be reasonably well-approximated by a
simple power-law behavior (Equation~\ref{eqn:mdot.time}), rising and
decaying from a characteristic maximum SFR with characteristic
half-life $t_{\rm burst}$ and a typical late-time power-law slope of
$\dot{M}_{\ast}\propto (t/t_{\rm burst})^{-\beta}$ where $\beta \sim
2.0-3.0$.

The characteristic burst mass $M_{\rm burst}$ is $\sim10\%$ of the
total spheroid mass, but it scales weakly with galaxy mass in a manner
similar to how disk galaxy gas fractions scale with their stellar
masses (expected if disks are the pre-merger progenitors of
spheroids), $M_{\rm burst}\approx
1/(1+[M_{\ast}/10^{9.15}\,\msun]^{0.4})$.  Detailed discussion of this
trend, and its consequences for the global structure and kinematics of
spheroids, are presented in \citet{hopkins:cusps.fp}. However, it
already makes it clear that bursts should not dominate the SFR
density.  They are a small fraction $\sim10\%$ of all stars in
spheroids (let alone all stars).  That does not mean that bursts are
unimportant, however.  It is clear that they control many of the
properties of galaxies \citep{cox:kinematics,robertson:fp,
naab:gas,onorbe:diss.fp.details,ciotti:dry.vs.wet.mergers,
jesseit:kinematics,jesseit:merger.rem.spin.vs.gas,
covington:diss.size.expectation, hopkins:cusps.ell,hopkins:cores}, and
they can account for the short-lived, highest-SFR systems in the
Universe. Moreover, the scatter in burst mass is significant,
$\sim0.3-0.4\,$dex, which is critical for explaining the most extreme
starbursts in the Universe, which require both large absolute galaxy
masses and gas fractions to reach the very high burst masses required.

The typical starburst timescale implied from the combination of
observed surface densities and the \citet{kennicutt98} law is a nearly
galaxy-mass independent $t_{\rm burst}\sim10^{8}\,$yr. Both the value
of this timescale, and the weak scaling with galaxy and/or burst mass,
agree well with the dynamical times in the central $\sim$kpc of
galaxies.  As above, though, there is considerable scatter of order
$\sim0.2\,$dex.  Given such a short starburst timescale, and the
relatively small total mass fractions involved in starbursts, it
naturally follows that starbursts will represent only a small fraction
of the star-forming galaxies at any stellar mass, at a particular
instant. Given that the average number of bursts per galaxy is not
large, the duty cycle of bursts should be $\sim t_{\rm burst}/t_{\rm
Hubble}$, or $\sim1-5\%$ from $z=0-3$.  Indeed, observations have
shown that most galaxies at these redshifts lie on a normal
star-forming sequence, without a large $\sim1\,\sigma$ scatter from
e.g.\ merger-induced bursts \citep{noeske:2007.sfh.part1,
noeske:sfh,papovich:ssfr,bell:morphology.vs.sfr}. The small duty cycle
here means that even if the burst SFR enhancement is large, it will
not violate these constraints (appearing only at the $\sim2-3\,\sigma$
level in the wings of the SFR distribution at a given mass).

These conclusions are supported by independent 
evidence from observational stellar population synthesis studies. 
Specific comparisons to the objects considered here, where 
available, are presented in detail in \citet{hopkins:cusps.ell,hopkins:cores} 
and \citep{foster:metallicity.gradients.prep}. It is well-established that 
constraints from abundances require the central portions of spheroids be formed 
in a similar, short timescale. And detailed decomposition of stellar populations into 
burst plus older stellar populations have yielded consistent results for the 
typical burst fractions and sizes \citep{titus:ssp.decomp,schweizer:7252,
reichardt:ssp.decomp,michard:ssp.decomp}. 

As should be expected from the generic behavior above, bursts peak at
SFRs of $\sim M_{\rm burst}/t_{\rm burst}$, which follows $M_{\rm
burst}$ (and hence total spheroid mass) in a close-to-linear relation.
The most massive local ellipticals -- especially those with total
stellar masses of $\gtrsim10^{12}\,\msun$ -- had extreme peak SFRs of
$>1000\,\msun\,{\rm yr^{-1}}$.  Thus, it is at least possible that
some local systems reached the highest SFRs inferred for massive,
high-redshift starburst galaxies
\citep{papovich:highz.sb.gal.timescales,
chapman:submm.lfs,walter:2009.hyperlirg.in.highz.qso.host}.  More
moderate, but still massive ellipticals with $M_{\ast}\sim
1-5\times10^{11}\,\msun$, reached a range of peak SFRs from $\sim
30-500\,\msun\,{\rm yr^{-1}}$, corresponding to their forming
fractions from $\sim5-20\%$ of their total masses in starbursts.
These match well with the observed SFRs in more typical, local and
$z\sim1$ starbursts (in $\sim L_{\ast}$ galaxies), which have been
specifically associated with mergers driving gas to galaxy centers,
and forming the appropriate nuclear mass concentrations to explain
e.g.\ elliptical kinematics, sizes, phase space densities, and
fundamental plane scalings \citep{cox:kinematics,
naab:gas,robertson:fp,jesseit:merger.rem.spin.vs.gas,hopkins:cusps.mergers,
hopkins:cusps.fp,hopkins:cusps.evol}.

We similarly quantify starburst spatial sizes as a function of their mass,
peak SFR, and time. Starburst sizes scale with starburst mass in a
similar fashion as the spheroid mass-size relation, but are smaller
than their host spheroids by a fraction similar to their mass
fraction.  The most massive starbursts reach half-SFR (i.e.\
half-light, in IR or mm wavelengths) size scales of $\sim
1-5\,$kpc. For typical spatial distributions, this implies total
spatial extents of strong emission up to $\sim10\,$kpc.  These size
scales also agree well with observations of the most massive
high-redshift starbursting systems \citep{younger:smg.sizes,
tacconi:smg.maximal.sb.sizes,
schinnerer:submm.merger.w.compact.mol.gas}, and of massive, compact
ellipticals formed at high redshift, believed to be the relics of such
starbursts \citep[with, at that time, little envelope of dissipationless,
low-density material yet accreted;][]{vandokkum:z2.sizes,
cimatti:highz.compact.gal.smgs,trujillo:compact.most.massive,
bezanson:massive.gal.cores.evol,hopkins:r.z.evol}.

For more typical, $\sim L_{\ast}$ starbursts, sizes range from
$\sim0.1-1\,$kpc, also similar to those observed in merging systems
\citep{scoville86,sanders:review,
hibbard.yun:excess.light,tacconi:ulirgs.sb.profiles,laine:toomre.sequence,
rj:profiles}. The size difference between these and the
most extreme objects follows from the much larger gas supply involved --
there is no dramatic difference in the relic starburst mass
distribution {\em shapes}.  Some claims have been made that the large
sizes of high-redshift starbursts could imply that they are not
scaled up analogues of local extreme starbursts; but we find here
that they correspond naturally.  Scaling up a starbursting system in
the starburst mass fraction will not preserve spatial size, but rather will
scale along the starburst size-mass relation here, which appears to be
smooth and continuous from the smallest starbursts with masses
$\sim10^{7}\,\msun$ to the largest with masses $>10^{11}\,\msun$.  The
origin of the size-mass scaling is of considerable
physical interest.  It has been proposed that in such starbursts the
Eddington limit from radiation pressure on dusty gas sets a universal
maximum central surface density, over all mass scales, from which this
follows \citep{hopkins:maximum.surface.densities}.  The important
constraint, from our analysis, is that there is no discontinuity, and
we provide the scaling relations that any such model must satisfy.

Combining these constraints with observational measurements of the
nuclear stellar population ages of these systems -- i.e.\ the
distribution of times when these bursts occurred -- we show that it is
possible to re-construct the dissipational burst contribution to the
distribution of SFRs and IR luminosity functions and luminosity
density of the Universe.  We show that the burst luminosity functions
agree well with the observed IR LFs at the brightest luminosities, at
redshifts $z\sim0-2$. At low luminosities, however, bursts are always
unimportant, as expected from their short duty cycles, noted above.
Although the burst luminosity functions rise with redshift, they
always represent low space densities, and the overall LF evolves
rapidly. As such, the transition luminosity above which bursts
dominate the IR LFs and SFR distributions increases with redshift from
the ULIRG threshold at $z\sim0$ to Hyper-LIRG thresholds at $z\sim2$.
This appears to agree well with recent estimates of the transition
between normal star formation and mergers, along the observed
luminosity functions.  Systematic morphological studies at low
redshifts \citep{sanders96:ulirgs.mergers} yield the conventional
wisdom that -- locally -- the brightest LIRGs and essentially all
ULIRGs are merging systems (see also references in
\S~\ref{sec:intro}).  At high redshifts, similar studies have now been
performed \citep[see e.g.][and references
therein]{tacconi:smg.mgr.lifetime.to.quiescent}.  They too find that
the brightest sources are almost exclusively mergers, but with a
transition point an order-of-magnitude higher in luminosity.  Other
morphological studies at intermediate redshifts $z\sim0.4-1.4$ have
reached similar conclusions \citep{bridge:merger.fractions}.

Integrating these luminosity functions, we find the burst contribution
to the SFR and IR luminosity densities of the Universe, and show that
it is small at all redshifts, rising from $\sim1-5\%$ at $z\sim0$ to a
roughly constant $\sim4-10\%$ at $z>1$.  This agrees well with recent
attempts to estimate the contribution to the SFR density at $z=0-1$
specifically from merger-induced starbursts, using either pair or
morphologically-selected samples \citep{jogee:merger.density.08,
robaina:2009.sf.fraction.from.mergers}.\footnote{Note that it is
important here to distinguish estimates of the SFR {\em induced} by
mergers, i.e.\ that above what some control population would exhibit,
from that simply in ongoing/identifiable mergers (since many criteria
identify mergers for a timescale $\sim$Gyr, much longer than the burst
timescale). For example, a merger fraction of $10\%$ would imply at
least $10\%$ of star formation in ongoing mergers, even if there were
no starbursts and those systems were only forming stars at the same
rate as they would in isolation.}  Given the completely independent
nature of the constraints, and significant uncertainties involved in
both, the agreement is good.  The small value is what is expected,
given our previous determination that the typical burst mass is just
$\sim10\%$ in $\sim L_{\ast}$ spheroids (the galaxies that dominate
the stellar mass density).  But it clearly rules out merger-induced
bursts driving the SFR density evolution of the Universe.

At the highest redshifts $z\gtrsim2$, we can put strict upper limits
on starburst intensities, based on the maximum stellar mass remaining
at high densities at $z=0$, and find some tension between these and
estimated number counts of sub-millimeter galaxies from
\citet{chapman:submm.lfs}.  This implies that some change may be
necessary in either the number counts themselves, the bolometric
corrections used to convert these observations to total IR
luminosities, or the stellar IMF used to convert between SFR and IR
luminosity.  However, the observations remain considerably uncertain,
with the bolometric corrections relying sensitively on assumed dust
temperatures, and the number counts subject to significant cosmic
variance \citep[see
e.g.][]{austermann:2009.aztec.submm.source.counts}.  More
observations, at new wavelengths and in larger, independent fields,
are needed to resolve these discrepancies.

We compare our constraints on these histories with recent predictions
from galaxy formation models and simulations, and find reasonably good
agreement.  Both exhibit similar tension with estimates of the
high-redshift, extremely luminous number densities. However, the
models are able to match the inferred number densities presented here
{\em without} any change to the stellar IMF or requiring other exotic
physics.  The systematic uncertainties in the models, especially at
high redshift, are large, however, so the empirical constraints
presented here provide a powerful new means of constraining the models
and their input parameters.

For example, if models are constrained via e.g.\ a halo occupation
approach or otherwise, so as to match the observed merger fractions of
galaxies as a function of redshift, then these numbers become
relatively large ($>10\%$) at redshifts $z\gtrsim2$ \citep[see
e.g.][]{bundy:merger.fraction.new,conselice:mgr.pairs.updated,
kartaltepe:pair.fractions,lin:mergers.by.type,bluck:highz.merger.fraction,
hopkins:merger.rates,jogee:merger.density.08,
bridge:merger.fraction.new.prep}. The most common assumption in many
analytic and semi-analytic models is that, in such a major merger, the
entire galaxy gas supply is channeled into the starburst. This,
however, coupled with the high observed merger fractions (and implied
merger rates), would lead to an over-prediction of the burst
contribution to the SFR density, relative to our constraints here. For
example, the predictions of such a model are given in
\citet{hopkins:merger.lfs}, and rise to $\sim20-50\%$ of the SFR
density at $z>1-2$.  Recently, however, \citet{hopkins:disk.survival}
have pointed out that the processes that lead to angular momentum loss
by the gas -- driving bursts in the first place -- rely on the stellar
component of merging systems, and so become less efficient as gas
fractions increase. Including these physics in the models (as the
current generation to which we compare does) leads to an asymptotic
maximum contribution to the SFR density similar to that found
here, while still giving a good match to observed merger fractions.

Despite the simple nature of the assumptions involved in our analysis,
we show in high-resolution hydrodynamic simulations that they work
well in allowing us to recover the star formation histories of bursts
(even where the simulated system is much more complex).  Especially in a
statistical sense, our numerical experiments suggest that this
approach is robust to a variety of detailed deviations from our
idealized assumptions.  Testing this (even in a few objects) via
high-resolution reconstruction of stellar populations in bursts,
directly from their observed spatially-resolved SEDs, would provide a
powerful test of this.  Insofar as we have compared with different
simulations and models, the dominant uncertainty in our analysis is
the nature -- in particular the slope, relevant for the extrapolation
to high gas surface densities -- of the \citet{kennicutt98} law.  We
have considered a range of slopes suggested by different
observations. Although they yield similar behavior at moderate and low
SFRs, after approximately $\sim10^{8}$\,yr from the beginning of a
burst, the predicted behavior at very early times in the bursts is
quite different.  A larger \citet{kennicutt98} relation index implies
higher peak SFRs and more sharply peaked starbursts, increasing the
predicted number counts of the most luminous sources and making it
relatively more easy for models to account for the most extreme
star-forming systems.  It therefore remains of considerable importance
for observations to probe both gas density measurements and full SFR
indicators in extreme systems, at low and high redshifts.

Finally, we note that our analysis is only possible because, as
indicated by basic dynamics, simulations, and observations of stellar
populations, starburst components of spheroids were formed in
dissipational (i.e.\ gas-rich, at their centers), rapid star forming
events. The dissipationless ``envelopes'' surrounding the central,
dense components in spheroids were not formed in such a manner (again
indicated by both their structural and kinematic properties, simulations of their
formation, and stellar population observations).  Rather, they
represent the debris of stars from disks which were formed pre-merger,
and assembled (and violently relaxed) dissipationlessly.  These stars
were formed over extended periods of time, with new gas accretion onto
the disk fueling new or continuous star formation 
\citep[e.g.][]{keres:cooling.clumps.from.broken.filaments}, 
as opposed to a
single massive inflow.  Especially in the most massive systems, they
are also assembled from multiple systems, via e.g.\ minor mergers
contributing tidal material to the extended ``wings'' well-known in
massive galaxies.  As such, there is not a straightforward means to
invert their surface stellar mass density profiles to obtain their star
formation history.  Such constraints will depend on other methods,
such as e.g.\ direct stellar population analysis. It should be borne
in mind that this represents $\sim90\%$ of the mass in most spheroids,
and so understanding star formation in disks remains critical to
understanding the origin of stellar populations in spheroids.

\acknowledgments 
We thank Chris Hayward and Josh Younger for helpful discussions throughout 
the development of this manuscript. 
Support for PFH was provided by the Miller Institute for Basic Research 
in Science, University of California Berkeley.
\\

\bibliography{/Users/phopkins/Documents/lars_galaxies/papers/ms}

\clearpage

\begin{footnotesize}
\begin{landscape}
\ctable[
  caption={{\normalsize Fits to IR LF of Bursts}\label{tbl:fit.lfs}},center
  ]{cccccccccccc}{
\tnote[]{Parameters are given for best fit to the redshift-dependent 
form of the IR LF (Equations~\ref{eqn:double.pwr.law}-\ref{eqn:param.z.evol}): 
$\Phi(z) = \phi_{\ast}/[(L/L_{\ast})^{\alpha} + (L/L_{\ast})^{\gamma}]$.\\ } 
\tnote[a-c]{Break luminosity $L_{\ast}$ as a function of redshift, 
per Equation~\ref{eqn:param.z.evol}: 
$\log{\{L_{\ast}/L_{\sun}\}} = L_{0} + L^{\prime}\,\xi + L^{\prime\prime}\,\xi^{2}$, 
where $\xi\equiv \log{(1+z)}$} 
\tnote[d-f]{Normalization $\phi_{\ast}$: 
$\log{\{\phi_{\ast}/{\rm Mpc^{-3}\,\log^{-1}{L_{\rm IR}}}\}} = 
\phi_{0} + \phi^{\prime}\,\xi + \phi^{\prime\prime}\,\xi^{2}$}
\tnote[g-i]{Faint-end slope $\alpha$: 
$\alpha = \alpha_{0} + \alpha^{\prime}\,\xi + \alpha^{\prime\prime}\,\xi^{2}$}
\tnote[j-l]{Bright-end slope $\gamma$: 
$\gamma = \gamma_{0} + \gamma^{\prime}\,\xi + \gamma^{\prime\prime}\,\xi^{2}$\\ }
\tnote[m]{Luminosity functions assuming the low-redshift fitted slope for the 
Kennicutt-Schmidt relation between star formation and gas surface 
density, $n_{K}=1.4$}
\tnote[n]{Same, assuming instead the steeper slope suggested by 
high-redshift observations, $n_{K}=1.7$\\ \ \\ 
All fits are based on $z=0-4$ results. Extrapolation beyond these 
redshifts should be performed with caution. Statistical uncertainties in 
the fitted parameters are given in parentheses.}
}{
\hline\hline
\multicolumn{1}{c}{$L_{0}(\pm \Delta L_{0})$\tmark[\ a]} &
\multicolumn{1}{c}{$L^{\prime}$\tmark[\ b]} & 
\multicolumn{1}{c}{$L^{\prime\prime}$\tmark[\ c]} & 
\multicolumn{1}{c}{$\phi_{0}$\tmark[\ d]} & 
\multicolumn{1}{c}{$\phi^{\prime}$\tmark[\ e]} & 
\multicolumn{1}{c}{$\phi^{\prime\prime}$\tmark[\ f]} & 
\multicolumn{1}{c}{$\alpha_{0}$\tmark[\ g]} & 
\multicolumn{1}{c}{$\alpha^{\prime}$\tmark[\ h]} & 
\multicolumn{1}{c}{$\alpha^{\prime\prime}$\tmark[\ i]} & 
\multicolumn{1}{c}{$\gamma_{0}$\tmark[\ j]} & 
\multicolumn{1}{c}{$\gamma^{\prime}$\tmark[\ k]}  & 
\multicolumn{1}{c}{$\gamma^{\prime\prime}$\tmark[\ l]}  \\
\hline\\
\multicolumn{12}{c}{Double Power-Law Fit - Standard Model\tmark[\ m]}\\
\hline
11.87(0.09) & 5.08(0.43) & -7.29(0.48)
& -6.29(0.17) & 3.76(0.85) & -2.48(0.94)
& 0.98(0.09) & 0.39(0.49) & -3.33(0.62)
& 2.68(0.16) & 7.13(0.82) & -11.38(0.94)\\
\hline
\\ 
\multicolumn{12}{c}{Double Power-Law Fit - Steeper Kennicutt-Schmidt Index\tmark[\ n]}\\
\hline
11.89(0.09) & 6.75(0.44) & -9.08(0.51)
& -6.27(0.15) & 1.47(0.78) & -0.15(0.90)
& 0.87(0.07) & 0.63(0.40) & -2.63(0.49)
& 2.29(0.12) & 8.11(0.70) & -12.32(0.84)\\
\hline\hline\\
}
\end{landscape}
\end{footnotesize}

\end{document}